\documentclass[
reprint,superscriptaddress,
nofootinbib,
amssymb,
aps,
pra,
]{revtex4-1}
\usepackage[dvipsnames]{xcolor}
\usepackage{soul}

\usepackage{graphicx}
\usepackage{dcolumn}
\usepackage{bm}
\usepackage[utf8]{inputenc}

\usepackage[hypertexnames=false,pdftex,colorlinks=true,urlcolor=blue,
            linkcolor=Blue,citecolor=RedViolet]{hyperref} 
\newcommand{\hyp}[1]{\hyperlink{#1}{\rm \color{black} #1}}
\usepackage{braket}
\usepackage{physics}
\usepackage{mathtools}
\usepackage{tikz-cd}
\usepackage{dsfont}
\usepackage{scalerel}
\usepackage{ulem}
\usepackage{cancel}

\usepackage{mathbbol}
\DeclareSymbolFontAlphabet{\amsmathbb}{AMSb}%
\newtheorem{theorem}{Theorem}
\newtheorem{Lemma}{Lemma}
\newtheorem{definition}{Definition}

\newcommand{\beq}{\begin{equation}}
\newcommand{\eeq}{\end{equation}}
\newcommand{\bse}{\begin{subequations}}
\newcommand{\ese}{\end{subequations}}
\newcommand{\bea}{\begin{eqnarray}}
\newcommand{\eea}{\end{eqnarray}}

\newcommand{\tp}{\intercal}
\def\ie{\textit{i.e., }}
\def\eg{\textit{e.g., }}

\def\opL{\mathcal{L}}

\def\Cset{\mathds{C}}
\def\Rset{\mathds{R}}

\def\x{\mathbf{x}}

\def\l{\mathbf{l}}
\def\xx{\mathbb{x}}
\def\uu{\mathbb{u}}
\def\yy{\mathbb{y}}
\def\zz{\mathbb{z}}
\def\lb{\bar{\mathbf{l}}}

\def\bxi{\boldsymbol{\xi}}
\def\bxip{\boldsymbol{\xi}^\prime}

\def\AA{\mathbb{A}}
\def\OO{\mathbb{O}}
\def\oo{\mathbb{o}}
\def\rr{\mathbb{r}}

\def\WW{\mathbb{W}}
\def\ss{\mathbb{s}}
\def\00{\mathbb{0}}
\def\gaga{\bar{\mathbb{g}}}

\def\kk{\mathbb{k}}
\def\kkth{\mathbb{k}^{\mathrm{th}}}

\def\CC{\mathbb{C}}
\def\ccb{\bar{\mathbb{c}}}
\def\DD{\mathbb{D}}
\def\dd{\mathbb{d}}
\def\jc{\mathbb{jc}}

\def\Sig{\mathbb{\Upsilon}}
\def\GM{\mathbb{\Gamma}}

\def\BB{\mathbb{B}}

\def\BBp{\mathbb{B}^\prime}
\def\In{\mathbb{1}}
\def\I2n{\mathbb{l}}
\def\I2n{\mathbb{I}}

\def\VV{\mathbb{V}}
\def\VVth{\mathbb{V}^{\mathrm{th}}}
\def\MM{\mathbb{M}}
\def\QQ{\mathbb{Q}}
\def\VVt{\mathbb{V}_{\! t}}
\def\UU{\mathbb{U}}
\def\UUt{\mathbb{U}_t}
\def\yy{\mathbb{y}}
\def\ww{\mathbb{w}}

\newcommand{\expv}[1]{\ensuremath{\langle{#1}\rangle}}
\newcommand{\JJJ}{\mathsf{J}}
\newcommand{\SSS}{\mathsf{S}}
\newcommand{\SSSp}{\tilde{\mathsf{S}}}
\newcommand{\SST}{\tilde{\mathsf{S}}}
\newcommand{\SSSth}{{\mathsf{S}}^{\mathrm{th}}}

\newcommand{\RRR}{\mathsf{R}}

\newcommand{\Lambt}{\mathbb{\Lambda}}

\def\hrho{\hat{\rho}}
\def\hrhog{\hat{\sigma}}
\def\hrhobgth{\hat{\bar{\sigma}}^{\mathrm{th}}}
\def\hrhobgt{\hat{\bar{\sigma}}}

\def\hH{\hat{H}}
\def\hHp{\hat{H}_{\mathrm{eff}}}
\def\hHo{\hat{H}_{\mathrm{OME}}}

\def\hL{\hat{L}}
\def\hXme{\hat{X}^{-}}
\def\hXma{\hat{X}^{+}}
\def\hX{\hat{X}}

\def\hq{\hat{q}}
\def\hp{\hat{p}}
\def\hx{\hat{x}}

\def\huno{\hat{1}}
\def\hbx{\hat{\mathbf{x}}}

\begin{document}

\title{Thermal equilibrium in Gaussian dynamical semigroups}

\author{Fabricio  Toscano} 
\email{toscano@if.ufrj.br  }
\affiliation{Instituto  de F\'isica,  Universidade  Federal do  Rio de  Janeiro,
  21941-972, Rio de Janeiro, Brazil}

\author{Fernando Nicacio}
\email{nicacio@if.ufrj.br} 
\affiliation{Instituto  de F\'isica,  
Universidade  Federal do  Rio de  Janeiro,
  21941-972, Rio de Janeiro, Brazil}
\affiliation{ Universit\"at Wien, NuHAG, Fakult\"at f\"ur Mathematik, 
              A-1090 Wien, Austria.                                   }
\date{\today}

\begin{abstract}
We characterize all Gaussian dynamical semigroups in continuous variables quantum systems of $n$-bosonic modes 
which have a thermal Gibbs state as a stationary solution.
This is performed through an explicit relation
between the diffusion and dissipation matrices, which characterize the semigroup dynamics, 
and the covariance matrix of the thermal equilibrium state.
We also show that Alicki's quantum detailed-balance condition, 
based on a Gelfand-Naimark-Segal inner product,  
allows the determination of the temperature dependence of the diffusion and dissipation matrices
and the identification of different Gaussian dynamical semigroups which share the same thermal equilibrium state.
\end{abstract}

\maketitle

\section{Introduction}

In modern Quantum Information Theory for continuous variable systems, {\it i.e}, 
systems described by $n$-bosonic modes, 
Gaussian channels are the standard models in most of the quantum communication 
protocols \cite{Holevo1999,Holevo2001,Holevo2002,Cerf-book2007,Caruso2008,Holevo2019}.
These channels are defined as those bosonic channels which transform Gaussian states 
into Gaussian states~\cite{Weedbrook2012,Adesso2014}. 
Further, these states have an exceptional role in quantum communication as, 
for example, they are optimal for the transmission of classical information 
through Gaussian bosonic quantum channels with additive capacity \cite{Wolf2006}.
\par 
The most general form of a one-parameter Gaussian channel of $n$-bosonic modes is 
a Gaussian Dynamical Semigroup (\hypertarget{GDS}{GDS})~\cite{Heinosaari2010,Toscano2021}, constituting thus  
the tool to describe the dynamics of all memoryless continuous-in-time 
Gaussian quantum channels. 
This is the reason why they are widely used to describe noisy quantum channels in 
continuous variable systems~\cite{Heinosaari2010, Giovannetti2010,DePalma2016}.
Further, \hyp{GDS}s are able to describe all processes which can 
formally be written as decomposition and production of noninteracting particles (or quasi-particles) 
which can be treated at least approximately as bosons~\cite{Alicki2007}. 
In this context, \hyp{GDS}s are known as quasi-free completely-positive semigroups~\cite{Vanheuverzwijn1978,Vanheuverzwijn1978errata,Demoen1979}, which happens, 
for example, in damped collective modes in deep inelastic collisions \cite{Sandulescu1987}.
\par
The dynamics of any Quantum Dynamical Semigroup (\hypertarget{QDS}{QDS}), 
not necessarily Gaussian, 
also known as Quantum Markov Semigroups \cite{Carlen2017},
is described by a master equation in the Lindblad form~\cite{Lindblad1976,Gorini1976}.
In an analogous way, a \hyp{GDS} verifies a Lindblad master equation 
where the unitary part of the evolution is set by a quadratic Hamiltonian 
which describes the $n$-bosonic modes. 
By a quadratic Hamiltonian, we mean one composed by products of any two canonical conjugate operators, positions and momenta, 
or the set of self-adjoint operators that constitutes a representation of the Heisenberg canonical commutation relations 
of the particular bosonic system considered \cite{Tarasov2008}.
Meanwhile, the non-unitary part is given by Lindblad operators corresponding to complex linear functions of positions and momenta, 
which simplifies the non-unitary dynamics to be described only in terms of two real $2n\times 2n$ matrices, 
the diffusion and dissipation matrices \cite{Nicacio2016,Toscano2021}. 
\par
Alternatively, the Weyl-Wigner representation~\cite{Ozorio1998} for the master equation of a \hyp{GDS}s 
can be employed and corresponds to a linear Fokker-Planck equation for the evolution of the Wigner function 
of the evolved states~\cite{Nicacio2010,Nicacio2016,Toscano2021}.
This is the quantum counterpart of the classical channels corresponding to Ornstein-Uhlenbeck processes, 
thus also known as Bose-Ornstein–Uhlenbeck semigroups~\cite{Carlen2017}, 
whose evolved probability distributions satisfy exactly the same Fokker-Planck equation as the \hyp{GDS}s.  
\par 
Notably when the \hyp{GDS} dynamics has a steady state, 
this will be a dynamically invariant Gaussian state which attracts, over long times, 
the evolution of any initial condition~\cite{Frigerio1977,Frigerio1978,Carmichael1999}. 
Thus, for a given quadratic Hamiltonian, 
the characterization of stationary situations in \hyp{GDS}s corresponds 
to find all the diffusion and dissipation matrices that allow a stationary state \cite{Toscano2021}. 
Of particular importance are the stationary states which also corresponds to 
thermal equilibrium states, characterized by a Gibbs state of a quadratic Hamiltonian. 
This is the main subject of our study.
\par 
Thermal equilibrium in \hyp{QDS}s in finite-dimensional quantum systems has its own well-established theory
based on the so called Quantum Detailed Balance Condition (\hypertarget{QDBC}{QDBC}), 
as first stated by Alicki~\cite{Alicki1976,Alicki2007}, see also \cite{Carlen2017}). 
Of note are the results in~\cite{Carlen2017} where the authors prove, using the \hyp{QDBC},
that the evolution governed by a \hyp{QDS} is a gradient flow, in a particular Riemannian metric on the set of states, 
for the relative entropy of a state with respect to a Gibbs state.
On the other hand, the study of thermal equilibrium in \hyp{GDS}s is scarce, 
which is particularly true for the multimode scenario;
a prominent exception is the extension of the results for \hyp{QDS}s 
in finite-dimensional quantum systems to the case of a one mode \hyp{GDS} performed in \cite{Carlen2017}.  
\par 
Here we fill this gap and give a complete characterization of all $n$-mode \hyp{GDS}s with a thermal equilibrium state.
To this aim, we employ the Fokker-Planck equation for the evolution of the Wigner function 
and show that the thermal probability current is always null in every phase space point. 
This enable us to conclude that \hyp{GDS}s with thermal equilibrium are characterized by a set of three commuting matrices.
Two of the them are the Hamiltonian matrices associated to the covariance matrix of the thermal state and
the diffusion matrix of the \hyp{GDS}. The third one is the skew-Hamiltonian matrix associated to 
the dissipation matrix of the \hyp{GDS}. 
By another side, 
this condition neglects that different \hyp{GDS}s, characterized by the diffusion and dissipation matrices,
may share the same thermal equilibrium state, 
as a consequence of the fact that the relation among that matrices does not set their temperature dependence.
To circumvent this, we show that these characterizations 
are possible by extending Alicki's \hyp{QDBC} to bosonic-mode systems.
\par
The extended \hyp{QDBC} leads to a master equation for a \hyp{GDS}s 
with the form of a Quantum Master Optical Equation (\hypertarget{QOME}{QOME})~\cite{GardinerZoller2000}  
and the temperature dependence of the diffusion and dissipation matrices is established
for this type of \hyp{GDS}s. 
From this temperature characterization, we establish that all \hyp{GDS} that leads to thermal equilibrium
satisfy a \hyp{QDBC} if we allow an arbitrary temperature dependence for coupling constants 
between the system and the environment.
Finally, we  discriminate the Hamiltonians corresponding to the unitary part of the dynamics of a \hyp{GDS} 
which allow the occurrence of the thermalization. Although,  
we show that the thermalization process itself is not affected by these Hamiltonians.
\par
The paper is organized as follows. In Sec.\ref{Sec:QuandBiGDSGau} we introduce the \hyp{GDS}s, 
their action on Gaussian states (Sec.\ref{GS}) and the Weyl-Wigner formalism to describe its dynamics (Sec.\ref{WFFPE}). 
In the introduction of Sec.\ref{TheEqui}, 
we establish the general time dependence of the first and second order moments in \hyp{GDS}s with stationary solutions. 
Then, in Sec.\ref{FormHamil} we set up the problem of having thermal equilibrium as stationary solutions 
and describe general properties that must be satisfied by \hyp{GDS}s with thermal equilibrium. 
Section \ref{condDDandCC} contains one of our main results: 
a theorem that characterizes all the \hyp{GDS}s with a thermal equilibrium state. 
The extension of the \hyp{QDBC} to $n$-bosonic mode systems is placed in Sec.~\ref{QDBC},
where five theorems are presented. These theorems completely characterize all \hyp{GDS}s 
satisfying the detailed balance.
In Section~\ref{secQOME}, 
we show that the master equation of \hyp{GDS}s satisfying the \hyp{QDBC} 
always corresponds to a \hyp{QOME}; this section finishes with a discussion about 
entanglement properties of its thermal equilibrium state solution.
We further explore the characterization of thermal equilibrium states in \hyp{GDS}s
that satisfy a \hyp{QDBC} in Sec.~\ref{ExamplesProperties}, where 
the temperature dependence of the diffusion and dissipation matrices is developed in Sec.~\ref{tempDDandCC},
the high and low temperature limits are described in Sec.~\ref{handltemplimit}, 
and in Sec.~\ref{diffusivereg} we explain the pure diffusive regime, 
where the stationary solution is lost. 
In Sec.~\ref{examples} we describe the necessary structure of a quadratic Hamiltonian, 
governing the unitary part of the \hyp{GDS}, that has a thermal equilibrium state. 
In this section we also clarify the role
of this Hamiltonian in the process of thermalization.
Finally, we summarize our findings in Sec.\ref{conclusions}.
Some auxiliary calculations and technical proofs are 
presented in the Appendixes~\ref{AppLema1},~\ref{AppDBimpliesStat},~\ref{AppGDScommod},
~\ref{Appinvaderiva},~\ref{AppcoolformGM}, and~\ref{AppComRel}.

\section{Gaussian Dynamical Semigroups}
\label{Sec:QuandBiGDSGau} 
%
In the Schr\"odinger picture, a \hyp{QDS} 
is ruled by the Lindblad master equation 
(\hypertarget{LME}{LME})~\cite{Lindblad1976,Gorini1976}
\begin{equation}
\label{eq:mastereq}
\dv{\hrho_t}{t} = \opL[\hrho_t] = \opL_{\mathrm{U}}[\hrho_t] + \opL_{\mathrm{NU}}[\hrho_t],
\end{equation}
where
\bse \label{eq:LULNU}
\bea \label{eq:LU}
&&
\opL_{\mathrm{U}}[ \,\cdot\, ] = -\frac{\imath}{  \hbar}[\hHp, \cdot\, ] \\
\label{eq:LNU}
&&
\opL_{\mathrm{NU}}[\, \cdot\, ]  = \frac{1}{2\hbar} \sum_{k=1}^K  \left( 2  \hL_k \cdot
\hL_k^\dag - \hL_k^\dag \hL_k \cdot  - \, \cdot \hL_k^\dag \hL_k \right)
\eea
\ese
are, respectively, the infinitesimal generators of the unitary and non-unitary parts of the evolution.
The operator $\hHp$ is the effective Hamiltonian of the system and 
$\hL_k (k = 1,...,K)$ are the Lindblad operators. 
%

With the help of the Hilbert-Schmidt inner product, $\expv{\hat A,\hat B}={\Tr}(\hat A^\dagger\hat B)$, 
the adjoint $\bar{\opL}$ of the superoperator $\opL$ is defined by
\begin{equation}\label{innerProd}
\expv{\opL[\hat A],\hat B}=\expv{\hat A,\bar{\opL}[\hat B]},  
\end{equation}
which enables us to write the Heisenberg picture of Eq.(\ref{eq:mastereq}) 
for an observable $\hat O_t$ \cite{Tarasov2008}: 
\begin{equation} \label{eq:mastereqHP}
\dv{\hat O_t}{t}=\bar{\opL}[\hat O_t]=\bar{\opL}_{\mathrm{U}}[\hat O_t]+\bar{\opL}_{\mathrm{NU}}[\hat O_t],
\end{equation}
where
\bse\label{opLUnitHR}
\bea \label{opLUnitHR1}
\bar{\opL}_{\mathrm{U}}[\, \cdot \, ] &=& \frac{\imath}{\hbar}[\hHp, \cdot\, ], \\
\label{opLUnitHR2}
\bar{\opL}_{\mathrm{NU}}[\cdot] &=& \frac{1}{2\hbar} \sum_{k=1}^K  \left( 2  \hL_k^\dag \cdot \hL_k 
                                    - \,\cdot \hL_k^\dag \hL_k - \hL_k^\dag \hL_k \cdot \, \right).  
\eea
\ese
%

Since $\opL$ is time-independent, the solution of~\eqref{eq:mastereq} is formally given by 
$\hrho_t = e^{t\opL} \hrho_0$, which is the evolution 
of an initial condition $\hat \rho_0$, 
and the set $\{\Lambda_t=e^{t\opL}\}_{t\geq0}$ is properly the \hyp{QDS} 
in the Schrödinger picture \cite{Breuer2002}.
The solution of~\eqref{eq:mastereqHP} is formally written as 
$\hat O_t = e^{t\bar{\opL}} \hat O_0$, which gives the evolution 
of an initial condition $\hat O_0$, where the set $\{\bar{\Lambda}_t=e^{t\bar{\opL}}\}_{t\geq0}$ 
is the Heisenberg picture version of the \hyp{QDS}.
\par
The kinematics of a system with $n$-bosonic modes 
is described by a $2n$-dimensional column vector of canonical operators,  
\beq
\label{opvecx}
\hbx = (\hq_1 , \ldots  , \hq_n , \hp_1 ,  \ldots , \hp_n)^\tp,
\eeq 
satisfying the canonical commutation relation 
$[ \hx_j ,  \hx_k ]  = \imath  \hbar \JJJ_{jk}  \huno$, where 
  \beq \label{matJ}
  \JJJ  = \begin{pmatrix}  0 &
  \In\\ -\In &  0\end{pmatrix}, \,\,\, \JJJ^{-1} =  -\JJJ =  \JJJ^\tp,  
  \eeq 
is a $2 n \times 2  n$ real antisymmetric symplectic matrix and 
$\In$ is the $n \times n$ identity matrix. 
\par
In a $n$-bosonic-mode system 
a Gaussian dynamical semigroup is a \hyp{QDS} with the Hamiltonian and Lindblad operators given by  
\bse\label{eq:quadlimham}
\bea
\label{Hquadratic}
\hHp &=&  \frac12 \hbx^\tp   \BBp \hbx  + \hbx^\tp  \JJJ  \bxip, \\
\label{LinearLindbOp}
\hL_k &=& \l_k^\tp  \JJJ \hbx, \,\,\, (k =1,..., K),
\eea
\ese
where $\BBp = (\BBp)^\tp$ is the Hessian matrix of the Hamiltonian, 
$\bxip$ is an $(2 n)$-dimensional real column vector,  
and the $\l_k$'s are $(2 n)$-dimensional complex column vectors.
In this case, according to \cite{Toscano2021}, 
the superoperator in~\eqref{eq:LNU} 
becomes\footnote{From now on, we will use $\tr$ 
to denote the trace of a matrix and $\Tr$ to denote the trace of an operator.}
\bea
\label{linearLME}
\opL_{\mathrm{NU}}^{\mathrm G}[\hrho_t]&=& \tr\!\left(
\CC\JJJ\pdv{}{\hbx}\, [\hbx^\tp \hrho_t] 
+\frac{1}{2}\DD\pdv{}{\hbx}\pdv{}{\hbx^\tp} \hrho_t \right), 
\eea
where $\pdv{}{\hbx}\cdot=\frac{\imath}{\hbar}[(\JJJ\hbx),\cdot]$ is a column vector operator 
whose components are $\pdv{}{x_k}\cdot=\frac{\imath}{\hbar}[(\JJJ\hbx)_k,\cdot]$ with $k=1,\ldots,2n$.
The matrices 
\bea \label{defMats}
\DD = \hbar\Re(\GM) \quad \text{and} \quad \CC = \Im(\GM),
\eea
are, respectively, the diffusion and dissipation matrices, 
both defined through the decoherence matrix 
\begin{equation}
\label{defGM}
\GM  =  \sum_{k=1}^K  \l_k  \l_k^\dag, 
\end{equation}
which is composed by the vectors in the Lindblad operators (\ref{LinearLindbOp}).
The adjoint generator $\bar{\opL}_{\mathrm{NU}}$ in \eqref{eq:mastereqHP}, 
using Eq.(\ref{opLUnitHR2}) for the present case, is 
\bea
\label{linearLMEadjoint}
\bar{\opL}_{\mathrm{NU}}^{\mathrm G}[\hat O_t]&=& \tr\!\left(
\CC\JJJ\hbx\pdv{}{\hbx^\tp} \hat O_t 
+\frac{1}{2}\DD\pdv{}{\hbx}\pdv{}{\hbx^\tp} \hat O_t \right).
\eea
\par
According to the definitions in (\ref{defMats}), 
$\DD = \DD^\tp \ge 0$ and $\CC = -\CC^\tp$. 
Note also that, according to (\ref{defGM}), $\GM \ge 0$, and thus
\begin{equation}
\label{DDplusiCC}
\hbar\GM=\DD+\imath  \hbar\CC \geq 0,
\end{equation}
which can be interpreted as a generalized fluctuation--dissipation
relation~\cite{WisemanMilburn2009}. 
For the next sections, a useful result concerning a relation between 
the matrices in (\ref{DDplusiCC}) is the following Lemma, 
which is proved in the Appendix~\ref{AppLema1}. 
\begin{Lemma} \label{Lemmadetc}
If $\det \CC\neq 0$ in (\ref{DDplusiCC}),
{\it i.e.}, $\CC$ is invertible, then 
both the diffusion and the decoherence matrices are invertible and 
strictly positive-definite, that is, $\DD>0$ and $\GM >0$.
\end{Lemma}

Two quantities of main importance for establishing the results of 
this work are the mean-value vector 
\beq \label{eq:mvdef}
\expval{\hbx}_t  = \Tr(\hrho_t \hbx) 
\eeq 
and the (dimensionless) covariance matrix  
\beq \label{eq:cvdef}
\VVt  = \frac{1}{2\hbar}  \Tr( \hrho_t(\hbx -  \expval{\hbx}_t)(\hbx -
  \expval{\hbx}_t)^\tp ). 
\eeq
Despite the evolution of the system state through a \hyp{GDS} can be 
analytically determined~\cite{Carmichael1999}, 
the description for the system behavior is improved 
when analyzing the evolution of these two moments. 
Taking the temporal derivative of above equations and 
using the \hyp{LME} in (\ref{eq:mastereq}) for the \hyp{GDS}, {\it i.e.}, 
with the operators in (\ref{eq:quadlimham}), 
the cyclicity of the trace together with 
the canonical commutation relation yield \cite{Nicacio2016}
\begin{equation}
\label{timeMean}
\dv{\expval{\hbx}_t}{t}  = 
\AA \expval{\hbx}_t - \bxi,
\end{equation}
and
\begin{equation}
\label{timeLyapunov}
\dv{\VVt}{t} =  \left(\AA \VVt+\VVt  \AA^{\!\tp} \right)+\frac{\DD}{ \hbar},
\end{equation}
where we defined the drift matrix 
\begin{equation}
\label{AA}
\AA = \JJJ \BBp-\CC\JJJ,
\end{equation}
for $\BBp$ from~\eqref{Hquadratic} and $\CC$ from~\eqref{defMats}.

By direct integration, 
the solutions of Eqs.\eqref{timeMean} and \eqref{timeLyapunov} are, respectively,   
\bse
\label{firstsecmoments}
\bea
\label{meanxtint}
\!\!\expval{\hbx}_t&=&e^{\AA t} \expval{\hbx}_0 - \int_{0}^{t}\! dt' e^{\AA t'} \bxi, \\
\!\!\VVt&=&e^{\AA t}\VV_0 e^{\AA{\!^\tp} t}+\frac{1}{\hbar}\int_{0}^{t}\!dt' \; e^{\AA (t-t')}\;\DD\; e^{\AA^{\!\tp} (t-t')} .
\label{variancet}
\eea
\ese
If the matrix $\AA$ is invertible, 
the integral in (\ref{meanxtint}) can be explicitly performed and this solution becomes
\beq
\label{meanxt}
\expval{\hbx}_t = e^{\AA t}(\expval{\hbx}_0-\AA^{-1}\bxi)+\AA^{-1}\bxi.
\eeq 

\subsection{Gaussian States}
\label{GS}
%
The formalism presented so far describes the action of a \hyp{GDS} on a generic quantum state. 
However, the term ``Gaussian'' in the acronym ``\hyp{GDS}'' refers to 
the fact that this kind of dynamics is a quantum channel 
that preserves the Gaussian character of an initial 
Gaussian state throughout the whole evolution.

The density operator $\hrhog_t$ of a Gaussian state can 
be expressed as~\cite{Holevo2019, Banchi2015}
\begin{equation}
\label{denopGS}
\hrhog_t  = \frac{ {e^{-\frac{1}{2\hbar}(\hbx  -  \expval{\hbx}_t)^\tp  \UUt  (\hbx  -
    \expval{\hbx}_t)}} }{ \sqrt{{\det}( \VVt  + \tfrac{\imath}{2} \JJJ)}}, 
\end{equation}%
which is completely determined only by the moments in Eqs.(\ref{eq:mvdef}) and (\ref{eq:cvdef}), 
where the mean-value is $\expval{\hbx}_t = \Tr(\hrhog_t \hbx)$ 
and the matrix $\UUt$ is given by
\beq
\label{defUUt}
\UUt
= 2  \imath \JJJ  \coth^{-1}\left( 2  \imath \VVt  \JJJ \right). 
\eeq   
Note that $\VVt  + \tfrac{\imath}{2} \JJJ \geq 0$ is the {\it bona fide} 
condition of a covariance matrix of a quantum state~\cite{Simon1994}, 
thus the determinant in the denominator~\eqref{denopGS} is never negative. 
When subjected to a \hyp{GDS}, the evolved state is like (\ref{denopGS}) with 
$\expval{\hbx}_t$ and $\VVt$ given in (\ref{firstsecmoments}). 

The relation between the matrices in (\ref{defUUt}) can be strengthened,   
which will be necessary for our future results. 
It is immediate from (\ref{defUUt}) that 
\begin{equation}
\label{UVJJVU}
\JJJ\UUt \VVt  = \VVt \UUt\JJJ\Leftrightarrow [\JJJ\UUt,\VVt\JJJ]=0; 
\end{equation}
however, we will prove this relation using well-known results in order to 
establish methods and notations for several future occasions.
First, we use the Williamson theorem \cite{Simon1994,Nicacio2021wil} 
which establishes that for every $2n\times 2n$ real symmetric and positive-definite matrix $\VVt$, 
{\it i.e.}, $\VVt^\tp=\VVt>0$, 
there exists a symplectic matrix $\SSS_t\in {\rm Sp}(2n,\mathds R)$ such that 
\beq
\SSS_t\VVt\SSS_t^\tp=\kk_t \oplus \kk_t, 
\eeq 
where $\kk_t =\mbox{diag}(\kappa_1(t),\ldots,\kappa_n(t))$ is the symplectic spectra of $\VVt$ 
and $\kappa_j(t)\geq 1/2 \, (j = 1,...,n)$ are the symplectic eigenvalues.
Next, we use the  following Lemma, 
also a consequence of the Williamson theorem.
\begin{Lemma} 
\label{LemmaJOdiag}
A Hamiltonian matrix\footnote{A real $2n\times 2n$ matrix $\MM$ is said Hamiltonian matrix iff $\JJJ\MM$ 
(or equivalently $\MM\JJJ$) is symmetric, where $\mathsf J$ is in (\ref{matJ}).} 
$\OO\JJJ$, where $\OO$ is symmetric and positive-definite and $\mathsf J$ is in (\ref{matJ}), 
is diagonalized by the similarity transformation 
\beq\label{eqlema21}
(\QQ\SSS)\,\OO\JJJ\,(\QQ\SSS)^{-1}=(\imath \oo)\oplus(-\imath\oo),
\eeq
where $\oo={\rm diag}(o_1,\ldots,o_n)$, 
$o_j>0$ ($j=1,\ldots,n$) are the symplectic eigenvalues 
of $\OO$ through $\SSS$, \ie $\SSS\OO\SSS^\tp=\oo\oplus\oo$, 
and $\QQ$ is the complex matrix \footnote{The matrix $\QQ$ is 
a member of the compact symplectig group 
${\rm Sp}(n):= {\rm Sp}(2n,\mathds C)\;\cap\;{\rm SU}(2n)$.}
\beq
\label{defQQ}
\QQ=\QQ^\tp = \frac{1}{\sqrt{2}}\begin{pmatrix} 
      \In&  -\imath \In \\
      -\imath  \In&   \In\\
   \end{pmatrix}
\eeq
such that $\QQ^\tp\JJJ\QQ=\JJJ$ and $\QQ^{-1}=\QQ^\dagger$.
Equivalently, we have
\beq\label{eqlema22}
(\QQ\SSS^{-\tp})\,\JJJ\OO\,(\QQ\SSS^{-\tp})^{-1}=(\imath \oo)\oplus(-\imath\oo).
\eeq
\end{Lemma}
\par
Returning to the proof of Eq.(\ref{UVJJVU}), the above Lemma can be used to diagonalize the matrix $\VVt\JJJ$, \ie 
\beq
\label{diagiVVJJ}
\QQ\SSS_t\,\VVt\JJJ(\QQ\SSS_t)^{-1}=(\imath \kk_t) \oplus (-\imath \kk_t) =: (\VVt\JJJ)_{\rm d}.
\eeq
As a useful notation, the diagonal matrix $(\VVt\JJJ)_{\rm d}$ defined above will be called 
the canonical form of $\VVt\JJJ$. 
From (\ref{defUUt}), we write $\JJJ\UUt= - 2\imath \, g(2\imath\VV_{\! t}\JJJ)$, 
where 
\beq
\label{defgx}
g(x) = 2\coth^{-1}(2x)>0
\eeq 
is a continuous function for $x > 1/2$.
Consequently, employing Eq.(\ref{diagiVVJJ}) and noting that $g(x)=-g(-x)$, we attain 
\bea
\label{JJJUUt}
\QQ\SSS_t\,\JJJ\UUt\,(\QQ\SSS_t)^{-1} = [\imath g(\kk_t)] \oplus [-\imath g(\kk_t)] =: (\JJJ\UUt)_{\rm d},
\eea
which is the canonical form of the Hamiltonian matrix $\JJJ\UUt$. 
Therefore, the matrix $(\QQ\SSS_t)^{-1}$ simultaneously diagonalizes the matrices 
$\VVt\JJJ$ and $\JJJ\UUt$ and they must commute, as we wanted to prove.

Noteworthy, the matrix $\UUt$ is positive-definite so the Wil\-liam\-son theorem can be applied.
The symplectic diagonalization can be easily inferred using~\eqref{eqlema22} with $\OO=\UUt$
and $\SSS^{-\tp}=\SSS_t$, 
so we get the canonical form in~\eqref{JJJUUt}.  
Therefore according to the hypothesis of the 
Lemma the matrix that diagonalizes symplectically $\UU_t$ is $\SSS_t^{-\tp}$, \ie
\beq
\SSS_t^{-\tp}\UU_t\SSS_t^{-1}=g(\kk_t)\oplus g(\kk_t),
\label{symplectdiagUU}
\eeq
where $g(\kk_t) =\mbox{diag}(g(\kappa_1(t)),\ldots,g(\kappa_n(t)))$ and  
$g(\kappa_j)> 0 \, (j = 1,...,n)$ are the symplectic eigenvalues of $\UUt$.
\par
From~\eqref{symplectdiagUU}, we see that
the matrix $\UUt$ is finite whenever $\kappa_j(t)>1/2$ for all $j=1,\ldots,n$ and for each fixed value of $t$.
In this case the  density operator $\hrhog_t$ in~\eqref{denopGS} corresponds to a full-rank mixed-state.
Still, the representation of Gaussian states as in~\eqref{denopGS} is also valid in  
the limit  $\kappa_j(t) \to 1/2 \, \forall j$, where both the matrix $\UUt$ and ${\det}( \VVt  + \tfrac{\imath}{2} \JJJ)$ diverge.
In this limit, we have $\hrhog_t\rightarrow \dyad{\Psi_t}{\Psi_t}$,  
where $\ket{\Psi_t}$ is an $n$-mode pure-Gaussian-state.   
When some but not all symplectic eigenvalues are such that 
$\kappa_j(t)=1/2$, the same divergences happen, and 
$\hrhog_t$ in~\eqref{denopGS} represents a rank-deficient mixed quantum-state in the limit  $\kappa_j(t) \to 1/2$.
In conclusion, 
$\hrhog_t$ in~\eqref{denopGS} is a valid representation of the density operator of any Gaussian state.


\subsection{Wigner Function and Fokker-Planck Equation}
\label{WFFPE}

In continuous variable systems, the following sets
$\{  \hat T_{\bxi}= e^{\frac{\imath}{\hbar}\hbx^\tp\JJJ\bxi} \,\, | \,\, 
\bxi \!\in\! \Rset^{2n} \}$ 
and 
$\{\hat R_{\x} = (4\pi\hbar)^{-n} \int d\bxi \,e^{\frac{\imath}{\hbar}\bxi^\tp\JJJ\x}\,\hat T_{\bxi} \,\, | \,\,  
\x \!\in\! \Rset^{2n}\}$ 
are basis of a vector space constituted by operators acting 
on the separable infinite-dimensional Hilbert space 
${\cal H}=\otimes_{j=1}^n{\cal H}_j$ of the $n$-mode bosonic system
\cite{Nicacio2021Weyl}. 
The elements of these sets are called translations and reflections, respectively, and refer  
to their action on the vector operator $\hbx$ in \eqref{opvecx}, corresponding to the Heisenberg picture, 
namely $\hat T_{\bxi}^\dagger\,\hbx \,\hat T_{\bxi}=\hbx+\bxi\hat 1$ and 
$\hat R_{\x}^\dagger\,\hbx\,\hat R_{\x}=-\hbx+2\x \hat 1$ \cite{Ozorio1998}.
Translation operators are unitary 
$\hat T^\dagger=\hat T_{\bxi}^{-1}=\hat T_{-\bxi}$ and reflection operators are unitary and Hermitian, \ie 
involutory operators, $\hat R_{\x}^2=\hat 1$. 
The operator $\hat T_{\bxi}$ is also known as Weyl operator and 
$\hat R_{\x}$ as Wigner operator~\cite{Tarasov2008}.
\par
When dealing with continuous variable systems, the existence of unbounded operators and
operators with continuous spectra may cause some mathematical difficulties. 
In particular, it is often difficult to find the algebra of operators 
that defines the domain of applicability of a given formalism.
In this work, we circumvent this difficulty 
by applying our formalism to the algebra of all operators with a  Weyl and Wigner representation.
This will be particularly important for the demonstration developed 
in Appendixes~\ref{AppDBimpliesStat} and~\ref{AppGDScommod}. 

%
%
The Weyl and Wigner representations of an operator $\hat A$ are, respectively, 
the Hilbert-Schimidt inner products 
$A(\bxi)= \langle \hat A, \hat T_{\bxi}\rangle$ and $A(\x)= \langle \hat A, \hat R_{\x}\rangle$,  
which are the coefficients of the expansion (also called symbols of $\hat A$)
in one of the mentioned bases through the Bochner integrals~\cite{deGosson2006}:
\beq
\hat A=\int \frac{d\bxi}{(2\pi\hbar)^n} \,A(\bxi)\,\hat T_{\bxi}=
\int \frac{d\x}{(\pi\hbar)^n} \,A(\x)\,\hat R_{\x}.  
\eeq
These facts are consequences of the orthogonality relations
$\langle \hat T_{\bxi}, \hat T_{\bxi^\prime} \rangle = 
2^n \langle \hat R_{\bxi},\hat R_{\bxi^\prime}\rangle = (2\pi \hbar)^n\delta(\bxi^\prime-\bxi)$.
In particular, 
the Wigner representation of the operator vector $\hbx$ in~\eqref{opvecx}, 
\beq
\label{vecx}
\x =\expval*{\hbx, \hat R_{\x}}= (q_1 , \ldots  , q_n , p_1 ,  \ldots , p_n)^\tp, 
\eeq 
is a real vector in phase-space. 
In the following, we will use the Wigner representation of the master equation for a \hyp{GDS}, 
which is nothing more than an alternative description of the system evolution. Through this representation we will establish our first result in Theorem~\ref{theoremvNUzero} in Section~\ref{condDDandCC}.
Also, this will be important
to prove the results in Appendixes~\ref{AppDBimpliesStat} and~\ref{AppGDScommod}.

The Wigner representation of the \hyp{LME} in~\eqref{eq:mastereq} for the operators in (\ref{eq:quadlimham}), 
see \cite{Carmichael1999,Nicacio2010,WisemanMilburn2009},   
is the Fokker-Planck equation \cite{Risken1996}
\beq
\dv{W(\x,t)}{t}=-\pdv{}{\x^\tp}\,[{\bf v }_{\mathrm{U}}(\x,t)+{\bf v }_{\mathrm{NU}}(\x,t)]
\label{FPEGDS}
\eeq
for the Wigner function\footnote{
The Wigner function is proportional to the Wigner symbol of the density operator, 
this proportionality guarantees the normalization of that function as a quasi-probability density, 
see \cite{Carmichael1999,Nicacio2021Weyl,Ozorio1998,deGosson2006}, for instance.}
$W(\x) =\frac{1}{(\pi \hbar)^n} \expval*{\hat \rho, \hat R_{\x}}$. 
In the above equation, we identify the Fokker-Planck current vectors:
\bse
\label{currentvec}
\bea
{\bf v }_{\mathrm{U}}(\x,t)&=&   
(\JJJ\BBp\;\x -\bxip) \, W({\bf x},t), \,\,\, 
\label{currentvecU}\\
{\bf v }_{\mathrm{NU}}(\x,t)&=& -
\tfrac{1}{2}\DD \frac{\partial }{\partial {\bf x}}W({\bf x},t) -\CC\JJJ\x\;,
\eea
\ese
corresponding, respectively, to the unitary (reversible) and non-unitary (irreversible) 
contributions to the evolution of $W(\bf x)$.
Note that the first term in Eq.\eqref{FPEGDS} is the Poisson bracket 
\beq
\label{PoissonB}
-\pdv{}{\x^\tp}\,[{\bf v }_{\mathrm{U}}(\x,t)]=
[H_{\mathrm{eff}},W(\x,t)]_{\rm cl}
\eeq
between the Wigner function and the Hamiltonian 
$H_{\mathrm{eff}}= \frac12 \x^\tp  \BBp \x +  \x^\tp \JJJ \bxip$, 
which is the Wigner symbol of \eqref{Hquadratic}.
\par
When the initial state is a Gaussian state, the evolved Wigner function
of $\hrhog_t$ in~\eqref{denopGS} is
\beq
\label{eq:WignerGauss}
W_{\rm G}(\x,t) =
\frac{\expval*{\hrhog_t,\hat R_{\x}}}{(\pi\hbar)^n}  = 
\frac{ e^{-\frac{1}{2\hbar}(\x  - \expval{\hbx}_t)^\tp  \VVt^{-1}  (\x  -
\expval{\hbx}_t)}}{(2  \pi  \hbar)^n \sqrt{{\det} \VVt}}, 
\eeq%
for $\expval{\hbx}_t$ and $\VVt$ in (\ref{firstsecmoments}). 
This function is a multivariate Gaussian distribution,   
so the Fokker-Planck currents in (\ref{currentvec}) 
are true probability currents given by
\bse \label{currentvecG}
\bea
\!\!\!\!\!\! {\bf v }^{\rm G}_{\mathrm{U}}(\x,t)&=&  (\JJJ\BBp\;\x -\bxip) \, W_{\rm G}({\bf x},t), \,\,\, 
\label{currentvecUG}\\
\!\!\!\!\!\! {\bf v }^{\rm G}_{\mathrm{NU}}(\x,t)&=&[\tfrac{1}{2\hbar}\DD\VV_{\! t}^{-1}(\x  - \expval{\hbx}_t)
 -\CC\JJJ\x ]W_{\rm G}(\x,t).
 \label{currentvecNU}
\eea
\ese

\section{Thermal equilibrium in GDS: general considerations}
\label{TheEqui}
%
For \hyp{GDS}s, if there exists a stationary state 
$\hrhog^{\rm S}$, it will be unique for any initial state
$\hrho_0$, \ie
$\lim_{t\to \infty}\hrho_t = \hrhog^{\rm S}$~\cite{Frigerio1977,Frigerio1978,Carmichael1999}.  
In particular, starting with an initial Gaussian state, 
the evolved state remains Gaussian throughout the whole evolution, 
therefore $\hrhog^{\rm S}$ is necessarily Gaussian.   

The first moments and covariance matrix of the stationary state can 
be determined through the asymptotic behavior of Eqs.\eqref{firstsecmoments} 
and for that we resort to the Lyapunov theory of stability \cite{Horn1991}. 
Note that the only way to erase all the information about any initial condition 
in Eqs.\eqref{firstsecmoments} is to admit a matrix $\AA$ with all of its eigenvalues with negative real part, 
which is the same as saying that $\AA$ is a {\it Hurwitz} matrix. 
Consequently, the covariance matrix of $\hrhog^{\rm S}$ is a solution of the Lyapunov equation
\bea \label{lyapnoveq1}
\dv{\VV^{\rm S}}{t}  = \left(\AA \VV^{\rm S} + \VV^{\rm S} \AA^{\!\tp} \right)+\frac{\DD}{ \hbar} = 0.
\eea
Recalling that $\VV^{\rm S}$ is strictly positive-definite $\VV^{\rm S} >0$ and together with the 
Hurwitz condition over $\AA$, the Lyapunov theorem \cite{Horn1991} sets that $\DD >0$. 
%
%
In this case, the linear dynamical system in (\ref{timeMean}) is said 
globally asymptotically stable (AS)\cite{Horn1991} 
and the solutions in (\ref{firstsecmoments}) attains the asymptotic values  
\bse
\bea
\label{meanxtAest}
\expval{\hbx}^{\rm S}&=&\AA^{-1}\bxip,\\
\VV^{\rm S}&=&\frac{1}{\hbar}\int_{0}^{\infty} dt \;e^{\AA t}\;\DD\; e^{\AA^\tp t},
\label{variancetAest}
\eea
\ese
where it is clear that any trace of the initial state disappears. 
%
\subsection{Gibbs States as Stationary States}\label{FormHamil}
 A stationary Gaussian state in a \hyp{GDS} corresponds to a thermal equilibrium state when 
$\hrhog^{\rm S}=\hrhog^{\rm{th}}$ is the Gibbs state 
\beq
\label{GibbsstateH}
\hrhog^{\mathrm{th}}=\frac{\hrhobgth}{{\cal Z}^{\mathrm{th}}}, \quad \hrhobgth =e^{-\beta\hH}, \quad
{\cal Z}^{\mathrm{th}}= {\Tr}( \hrhobgth), 
\eeq
where $\beta$ is the ``inverse temperature'' and $\hH=\hHp$ is the quadratic Hamiltonian
in (\ref{Hquadratic}) or, more generically, another quadratic Hamiltonian such that $[\hH,\hHp]=0$.
In the following we will establish necessary conditions over the quadratic Hamiltonians, $\hHp$ and $\hH$, 
which allows a \hyp{GDS} to have an equilibrium thermal state.

Without loss of generality, 
we can set the origin of the  phase-space coordinates $\x$ in~\eqref{vecx} 
such $\expval{\hbx}^{\mathrm{th}}=\Tr(\hrhog^{\mathrm{th}}\hbx)=-\BB^{-1}\JJJ\bxi=0$, where
the vector $\bxi$ is associated with a possible linear term of $\hH$.
Therefore, the quadratic Hamiltonian $\hH$ can be chosen as
\beq
\label{Hform}
\hH=\tfrac{1}{2}\hbx^\tp\BB\hbx,
\eeq
\ie with $\bxi=0$.
The Hessian matrix of the Hamiltonian has to be positive definite, $\BB>0$, 
in order to fulfill the normalization condition 
${\Tr}(\hrhog^{\mathrm{th}})=1$ \cite{Nicacio2021Weyl}. 
From $\hHp$ in~\eqref{Hquadratic} and $\hH$ in~\eqref{Hform}, we have 
\bea \label{commHam}
[\hat H,\hat H_{\mathrm{eff}}]& =& 
-\tfrac{\imath\hbar}{2}\hbx^\tp\JJJ[\JJJ\BB,\JJJ\BB^\prime]\hbx-\imath \hbar(\bxi^\prime)^\tp\BB\hbx=0\nonumber\\
&\Leftrightarrow&
[\JJJ\BB,\JJJ\BB^\prime]=0\;\;\text{and}\;\;\bxi^\prime=0.
\eea
Therefore, the Hamiltonian 
of the free evolution of a \hyp{GDS} with an equilibrium thermal state must also be of the form
\beq
\label{Hformp}
\hHp=\tfrac12 \hbx^\tp   \BBp \hbx.
\eeq

\par
Comparing the general form of the density operator of a Guassian state, Eq.\eqref{denopGS}, with the 
thermal state $\hrhog^{\mathrm{th}}$ in~\eqref{GibbsstateH} with $\hH$ in~\eqref{Hform}, we arrive to
\beq
\label{UUth}
\UU^{\mathrm{th}}=\hbar \beta \BB.
\eeq
Applying condition \eqref{UVJJVU}, we get
\beq
\label{BBVVthJJJ}
\JJJ\BB\VVth=\VVth\BB\JJJ\Leftrightarrow [\JJJ\BB,\VVth\JJJ]=0,
\eeq
where $\VVth$ is the covariance matrix of the Gibbs Gaussian state in (\ref{GibbsstateH}).
Using \eqref{defUUt}, it is clear that $\VVth\JJJ$ is a function of $\JJJ\BB$, 
{\it viz}, 
\beq
\label{VVthJfuncofJB}
\VVth\JJJ=-\frac{\imath}{2}\coth\left(\frac{\imath\hbar \beta}{2}\JJJ\BB\right). 
\eeq
Taking into account the relation $[\JJJ\BB,\JJJ\BBp]=0$ from (\ref{commHam}) and that 
$\VVth\JJJ$ is a function of $\JJJ\BB$,
it is also true that
\beq
\label{BBpVVthJJJ}
[\JJJ\BBp,\VVth\JJJ]=0 \Leftrightarrow \JJJ\BBp\VVth=\VVth\BBp\JJJ.
\eeq
\par

According to Williamson theorem it is possible to find a symplectic matrix $\SSSth$ such that 
\bea\label{VVthsymdiag}
&&(\SSSth) \VVth (\SSSth)^{\tp} =\kkth\oplus\kkth,  
\eea
and using Eq.\eqref{eqlema21} of Lemma~\ref{LemmaJOdiag}, the matrix $(\QQ\SSSth)$ diagonalizes $\VVth\JJJ$:  
\beq
\label{diagiVVth}
\QQ\SSSth\,\VVth\JJJ(\QQ\SSSth)^{-1}=(\imath \kkth) \oplus (-\imath \kkth). 
\eeq
However, due to \eqref{VVthJfuncofJB}, the same matrix $(\QQ\SSSth)$ also diagonalizes $\JJJ\BB$.  
Using Eq.\eqref{symplectdiagUU}  one realizes that ${({\SSSth})}^{-\tp}$ 
is the symplectic matrix that diagonalizes $\BB$, \ie
\beq
\label{diagsymplecBB}
{({\SSSth})}^{-\tp} \BB (\SSSth)^{-1} = \ww\oplus\ww, \,\,\, \ww=\text{diag}(\omega_1,\ldots,\omega_n), 
\eeq
where $\omega_j>0$ ($j=1,\ldots,n$) are the symplectic eigenvalues 
of the Hessian matrix $\BB>0$ in~\eqref{Hform}. 
It is also possible to define, again according to Lemma~\ref{LemmaJOdiag},
the canonical form of $\JJJ\BB$:
\beq
\label{SpecJB}
(\JJJ\BB)_{\rm d} = (\QQ\SSSth)\, \JJJ\BB \,(\QQ\SSSth)^{-1} = 
(\imath\ww)\oplus (-\imath\ww).
\eeq
Finally, from (\ref{VVthJfuncofJB}), 
the relations between the symplectic spectra of $\VVth$ and $\BB$ is
\bea
\label{VVthsymdiag2}
\kkth = \frac{1}{2}\coth\left(\frac{\hbar \beta\ww}{2}\right), 
\eea
or equivalently $\ww=\frac{1}{\hbar\beta}g(\kkth)$, see Eq.(\ref{defgx}).

For a phase space described by $\x$ in (\ref{vecx}), 
the classical counterpart of the Hamiltonian $\hat H$ in~\eqref{Hform} coincides with 
its Wigner symbol, \ie
\beq\label{quadclassham}
H=\tfrac{1}{2}\x^\tp\BB\x= \langle \hat H, \hat R_{\x}\rangle
\eeq 
and the solution of the Hamilton equation
$\dot{\x}=\JJJ\pdv{H}{\x}$ is given by $\x(t)=\SST_t\x(0)$ with 
\begin{equation}\label{SSStdef}
\SST_t = e^{\JJJ\BB t} = (\QQ\SSSth)^{-1} \, e^{(\JJJ\BB)_{\rm d} t} \,(\QQ\SSSth)     
\end{equation}
and $(\JJJ\BB)_{\rm d}$ in (\ref{SpecJB}). 
The matrix $\SST_t$ defined above will be important in Section~\ref{condDDandCC}, 
but here it is worth to note that it generates a Hamiltonian flow 
around the elliptical fixed point $\x=0$ and that, from (\ref{diagsymplecBB}), 
$\omega_j$ are the eigenfrequencies of the Hamiltonian (\ref{quadclassham}). 
This is a direct consequence of the positive-definiteness of $\BB$ 
and, by this reason, we call positive-elliptic
all the Hamiltonians in~\eqref{Hform} with $\BB >0$. 
Consequently, a positive-elliptic Hamiltonian in~\eqref{Hform} is a necessary 
condition for a \hyp{GDS} to have an $n$-mode equilibrium thermal-state, 
since it is necessary for the convergence of the partition function 
${\cal Z}^{\mathrm{th}}$ in \eqref{GibbsstateH}.

Up to this point we were describing some properties 
of thermal states associated to quadratic Hamiltonians. 
In the next section, we will show the conditions over the diffusion and dissipation matrices $\DD$ and $\CC$, respectively,
which define a \hyp{GDS} with a thermal equilibrium state.

\subsection{Diffusion and dissipation matrices 
for thermal equilibrium}\label{condDDandCC}
%
The Lyapunov equation \eqref{lyapnoveq1} for the covariance matrix of a \hyp{GDS} thermal-equilibrium-state $\VVth$, 
where $\AA$ is given in (\ref{AA}),  
attains a simpler form through condition \eqref{BBpVVthJJJ}: 
\begin{equation}
\label{LyapunoveqVth}
 \CC \JJJ  \VVth+ \VVth \JJJ \CC = \frac{\DD}{\hbar},
\end{equation}
whose unique formal solution, see Eq.(\ref{variancetAest}), is 
\beq
\label{VVthinte}
\VVth=\frac{1}{\hbar}\int_{0}^{\infty} dt \;e^{-\CC\JJJ t}\;\DD\; e^{-\JJJ\CC t},
\eeq
where according to  Lyapunov theorem \cite{Horn1991}, 
the matrix $-\CC\JJJ$ must be {\it Hurwitz}, 
since $\VVth > 0$ and $\DD >0$. 
Therefore, the dissipation matrix $\CC$ must be invertible and according to Lemma~\ref{Lemmadetc}, 
$\DD>0$ and $\GM>0$, \ie the diffusion and decoherence matrices, 
in~\eqref{defMats} and~\eqref{defGM} respectively, 
of a \hyp{QDS} with a thermal equilibrium state must be positive definite.
In the following we show that an explicit solution of the integral~\eqref{VVthinte} 
can be obtained from the stationary condition over the  Fokker-Planck equation \eqref{FPEGDS} 
corresponding to a \hyp{QDS} with a thermal equilibrium state.
\par
When a \hyp{GDS} has a stationary state, 
this is a Gaussian state $\hrhog^s$ and from the Fokker-Planck equation \eqref{FPEGDS}, 
we obtain the condition
\beq
\label{divcurrzero}
-\pdv{}{\x^\tp}\,[{\bf v }^G_{\mathrm{U}}(\x)+{\bf v }^G_{\mathrm{NU}}(\x)]=0,
\eeq
with ${\bf v }^G_{\mathrm{U}}(\x)$ and ${\bf v }^G_{\mathrm{NU}}(\x)$ in Eqs.\eqref{currentvecG}. 
If this stationary state is a thermal equilibrium state, 
$\hrhog^s=\hrhog^{\mathrm{th}}$, 
condition~\eqref{divcurrzero} simplifies to
\beq
\label{divcurrzeroth}
-\pdv{}{\x^\tp}\,[{\bf v }^{\mathrm{th}}_{\mathrm{NU}}(\x)]=0, 
\eeq
since, due to Eq.\eqref{PoissonB}, one has 
\bea
[H_{\mathrm{eff}}(\x),W^{\mathrm{th}}(\x)]_{\mathrm{cl}} &=&\tr\left(\JJJ\BBp\x\x^\tp(\VVth)^{-1}\right)=0,
\eea
where we employed Eq.\eqref{BBpVVthJJJ} and the fact that $\tr(\AA)=\tr(\AA^\tp)$ for any matrix $\AA$. 
Now we can establish the following theorem that characterizes a \hyp{QDS} 
with a thermal equilibrium state:
\begin{theorem}
\label{theoremvNUzero}
A \hyp{QDS} has a thermal equilibrium state iff 
\beq
\label{currvthzero}
{\bf v }^{\mathrm{th}}_{\mathrm{NU}}(\x)=0.  
\eeq
The covariance matrix of such state is given by 
\beq
\label{Vthposta}
\JJJ \VVth=\frac{1}{2\hbar}\JJJ\DD(\JJJ\CC)^{-1}, 
\eeq
where 
\beq
\label{JDcommuteJC}
[\JJJ\DD,\JJJ\CC]=0.
\eeq 
\end{theorem}
In order to prove our theorem, 
we use the Divergence Theorem\footnote{See, for example, Appendix A of \cite{Toranzo2017}.}
and~\eqref{divcurrzeroth}, 
both enable us to relate the divergence of the vector field ${\bf v }^{\mathrm{th}}_{\mathrm{NU}}(\x)$ 
with the flux through the boundary $\partial \Omega$ of the region $\Omega\in \Rset^{2n}$,  
\beq
\label{DTovervNU}
\int_{\partial \Omega}\,
{\bf n}^\tp {\bf v }^{\mathrm{th}}_{\mathrm{NU}}(\x)\,ds=
\int_{\Omega} \pdv{}{\x^\tp}\,[{\bf v }^{\mathrm{th}}_{\mathrm{NU}}(\x)] \,d\x^{2n}=0,
\eeq
where ${\bf n}$ is the $2n-$dimensional real vector normal 
to the surface $\partial \Omega$. 
Since $\Omega$ has arbitrary volume, 
the necessary and sufficient condition in (\ref{currvthzero}) is proved. 
From 
~\eqref{currentvecNU}, 
${\bf v }^{\mathrm{th}}_{\mathrm{NU}}(\x)=\left(\frac{1}{2\hbar}\DD(\VVth)^{-1}-\CC\JJJ\right)\x \,W^{\mathrm{th}}(\x)=0$ 
for any $\x$, ending up with Eq.\eqref{Vthposta}. 
The relation in Eq.\eqref{JDcommuteJC} follows from the requirement $\VVth=(\VVth)^\tp$. 
Note that, if $[\JJJ\DD,\JJJ\CC]=0$
then $\DD\;e^{-\JJJ\CC t}=e^{-\CC\JJJ t}\;\DD$ and the integration in~\eqref{VVthinte} 
can be explicitly performed to obtain exactly the expression in~\eqref{Vthposta}.

\par
It is worth to note that the covariance matrix $\VVth$ of the thermal equilibrium state $\hrhog^{\mathrm{th}}$ in~\eqref{GibbsstateH} 
is completely determined by the Hessian $\BB$ of the Hamiltonian $\hat H$ in~\eqref{Hform}.
This is shown in Eqs.\eqref{VVthsymdiag} and~\eqref{VVthsymdiag2}, 
where ${({\SSSth})}^{-\tp}$ symplectically diagonalizes $\BB$,
whose symplectic spectrum is contained in the diagonal matrix $\ww$. 
So, the expression in Eq.\eqref{Vthposta} simply establishes the connection between the fixed matrix $\VVth$ 
and the dynamics of the \hyp{GDS}, that is, the one determined by the matrices $\CC$ and $\DD$, which at the end  
determines $\hrhog^{\mathrm{th}}$ as an equilibrium state.
However, the Lyapunov equation~\eqref{LyapunoveqVth} has common solutions \cite{Bialas2015}, 
\ie there are different matrices $\DD$ and $\CC$ which are able to give the same covariance matrix $\VVth$ in~\eqref{Vthposta}. 
Each pair $(\DD,\CC)$ corresponds to a different \hyp{GDS} which has as steady state the same thermal state $\hrhog^{\mathrm{th}}$. 
In the next section we show that the \hyp{QDBC} determines common solutions.
\par
Also, according to Theorem~\ref{theoremvNUzero}, a \hyp{QDS} has a thermal equilibrium state 
iff the set $\{\JJJ\VVth,\JJJ\DD,\JJJ\CC\}$ is a commuting set of matrices. 
Therefore, there is a matrix that simultaneously diagonalizes the three matrices $\JJJ\VVth$, $\JJJ\DD$, and $\JJJ\CC$ 
(see Appendix~\ref{AppComRel}). 
Using Lemma~\ref{LemmaJOdiag}, the Hamiltonian matrix 
$\JJJ\VVth$ is diagonalized by 
$(\QQ\SSSth)^\dagger$ 
so, for a \hyp{GDS} with a thermal equilibrium state, we can always write 
\beq
  \label{Comsetdiag1}
\JJJ\VVth=(\QQ\SSSth)^\dagger\;(\JJJ\VVth)_{\rm d}\;((\QQ\SSSth)^\dagger)^{-1},
\eeq
with $(\JJJ\VVth)_{\rm d}= \imath\kkth \oplus  \left(-\imath\kkth\right)$ and $\kkth$ in~\eqref{VVthsymdiag2}.
Equivalently, applying the same Lemma to the diagonalization of $\JJJ\DD$ we arrive at 
\beq
  \label{Comsetdiag2}
\JJJ\DD = (\QQ \SSSth)^\dagger \; (\JJJ\DD)_{\rm d} \; ((\QQ\SSSth)^\dagger)^{-1},
\eeq%
where $(\JJJ\DD)_{\rm d}=(\imath \hbar \dd)\oplus(-\imath\hbar\dd)$ with $\hbar \dd$ the $\Rset^{n\times n}$ 
diagonal matrix with the symplectic spectrum of $\DD$ in its diagonal%
\footnote{It is worth to note that $\SSSth$ simultaneously diagonalizes simplectically  
$\VVth$ and $\DD$.}. 
Because $\JJJ\CC$ is a skew-Hamiltonian matrix
\footnote{A real $2n\times 2n$ matrix $\MM$ is said a skew-Hamiltonian matrix iff $\JJJ\MM$ (or equivalently $\MM\JJJ$) is a skew-symmetric matrix.},  its eigenvalues are real and with at least multiplicity equal to two~\cite{Fassbender1999}. So we can write 
\beq
\label{Comsetdiag3}
\JJJ\CC=(\QQ\SSSth)^\dagger\;(\JJJ\CC)_{\rm d}\;((\QQ\SSSth)^\dagger)^{-1},
\eeq
where $(\JJJ\CC)_{\rm d}=\jc\oplus \jc$ with $\jc$ a $\Rset^{n\times n}$ diagonal matrix.
Therefore, using~\eqref{Vthposta}  we arrive to 
$\frac{1}{2\hbar}(\JJJ\DD)_{\rm d}(\JJJ\CC)_{\rm d}^{-1}=(\JJJ\VVth)_{\rm d}=(\imath\kkth)\oplus(-\imath\kkth)$, or equivalently to
\beq
\label{cucamuruca}
\dd(\jc)^{-1}=2\kkth=\coth\left(\frac{\hbar \beta\ww}{2}\right)=\frac{\In+e^{-\hbar\beta\ww}}{\In-e^{-\hbar\beta\ww}}.
\eeq
However, this relation says nothing about the dependence on $\beta$, the inverse temperature,  
of the matrices $\dd$ and $\jc$ composed by the eigenvalues of the matrices $\JJJ\DD$ and $\JJJ\CC$, respectively. 
In the next section we will show that a \hyp{QDBC} allows
the determination of this dependence.

%

\section{ GDS\lowercase{s} satisfying a detailed balance condition}
\label{QDBC}
The notion of detailed balance is the principle governing the way thermal equilibrium is attained 
by classical Markov processes \cite{Risken1996}. 
It has several different quantum versions (see \cite{Carlen2017} and references therein)  
and in the context of \hyp{QDS} for finite-dimensional systems, 
the one due to Alicki~\cite{Alicki1976} stands out because it allows the extension 
of time-reversal invariance 
of classical equilibrium to the quantum realm \cite{Carlen2017}.

Inspired by the classical case in Markov processes, 
where the time-reversal invariance of transition probabilities 
is related to a particular definition of an inner product, 
Alicki's definition for quantum detailed balance is based on the 
$\hat{\bar{\rho}}$-Gelfand-Naimark-Segal ($\hat{\bar{\rho}}$-\hypertarget{GNS}{GNS}) 
inner product in finite dimension Hilbert spaces:
\beq
\label{GNSinnerprod}
\expv{\hat A,\hat B}_\text{GNS}=\Tr(\hat{\bar{\rho}}\, \hat A^\dagger \hat B),
\eeq
where $\hat{\bar{\rho}}$ is a positive operator\footnote{The operator $\hat{\bar{\rho}}$ in Alicki's 
work~\cite{Alicki1976} is a full rank density operator in finite dimensional systems. 
However, it is more convenient to extend 
the definition of a $\hat{\bar{\rho}}$-\hyp{GNS} inner product for unnormalized density operators $\hat{\bar{\rho}}$ 
and, in particular, to unnormalized Gibbs states like 
$\hrhobgth $ in~\eqref{GibbsstateH}, see \cite{Carlen2017}.}, 
$\hat{\bar{\rho}} = \hat{\bar{\rho}}^\dagger>0$, and 
the operators $\hat A$ and $\hat B$ belong to a finite-dimensional C$^*$-algebra. 
Thus, Alicki's  
\hyp{QDBC} relies on the notion of self-adjointness  
with respect to the $\hat{\bar{\rho}}^{\rm th}$-\hyp{GNS} inner product,
where $\hat{\bar{\rho}}^{\rm th}=e^{-\beta \hat H}$ is an unnormalized Gibbs state with Hamiltonian $\hat H$.  
A superoperator $\Lambda$ is said to be self-adjoint 
with respect to the $\hat{\bar{\rho}}$-\hyp{GNS} inner product if
\bea
\expval{\Lambda[\hat A],\hat B}_\text{GNS}=\expval{\hat A,\Lambda[\hat B]}_\text{GNS},
\label{selfadjGNS}
\eea
for any operators $\hat A$ and $\hat B$ in the C$^*$-algebra.
\par
The extension of Alicki's approach for continuous variable systems relies on a 
definition for the set of operators where the \hyp{GNS}-inner-product is well defined.
In this regard, we consider operators acting on the separable Hilbert space of $n$-bosonic modes, 
${\cal H}=\otimes_{j=1}^n{\cal H}_j$, of infinite dimension. 
In our case, the $\hrhobgth$-\hyp{GNS} inner products are computed with $\hrhobgth$ in~\eqref{GibbsstateH} and $\hH$
being the quadratic Hamiltonian in~\eqref{Hform}. 
Also, regardless of whether the operators $\hat A$ and $\hat B$ are bounded or unbounded, having continuous spectra or not,
the domain of applicability of the $\hrhobgth$-\hyp{GNS} inner product in~\eqref{GNSinnerprod} with $\hat{\bar{\rho}} = \hrhobgth$
is over all operators such that the trace on this formula is finite\footnote{The existence of the trace in~\eqref{GNSinnerprod} can be checked, 
for example, using $\expv{\hat A,\hat B}_\text{GNS}=\int_{\Rset^{2n}}\,d\bxi\,\bar{\sigma}^{\rm th}(\bxi)(A^\dagger B)(\bxi)=\int_{\Rset^{2n}}\,d\x\,\bar{\sigma}^{\rm th}(\x)(A^\dagger B)(\x)$, where in the integrands we have the Weyl and Wigner symbols of 
$\hrhobgth$ and $\hat A^\dagger \hat B$, respectively. Notwithstanding, any other representation that could be more convenient can be used.}.
\par

For the definition of the \hyp{QDBC} in the context of \hyp{GDS}s, 
we recall the notation of Sec.\ref{Sec:QuandBiGDSGau}, 
where $\bar{\Lambda}_t=e^{t\bar{\opL}}$ with $\bar{\opL}$ in~\eqref{eq:mastereqHP} represents the superoperator that generates 
a \hyp{GDS} in the Heisenberg picture and ${\Lambda}_t=e^{t\opL}$ with $\opL$ in~\eqref{eq:mastereq},  the one in the Schrodinger picture. \\

\begin{definition}
Consider the \hyp{GDS} $\{\bar{\Lambda}_t=e^{t\bar{\opL}}\}_{t\geq0}$ 
with the infinitesimal generator $\bar{\opL}=\bar{\opL}_{\mathrm{U}}+\bar{\opL}_{\mathrm{NU}}^{\mathrm G}$, 
where $\bar{\opL}_{\mathrm{U}}$ is defined in~\eqref{opLUnitHR1} for the quadratic Hamiltonian 
in~\eqref{Hformp} and $\bar{\opL}_{\mathrm{NU}}^{\mathrm G}$ is defined in~\eqref{linearLMEadjoint}. 
This \hyp{GDS} satisfies the \hyp{QDBC} with respect to $\hrhobgth$ in~\eqref{GibbsstateH}
if $\bar{\Lambda}_t$ is self-adjoint with respect to the $\hrhobgth$-{\hyp{GNS}} inner product for all $t$.
In such case, we say that $\bar{\Lambda}_t$ satisfies a $\hrhobgth$-{\hypertarget{DBC}{DBC}}.
\end{definition}

The connection between the \hyp{QDBC} and a steady state of the \hyp{GDS} is in the following theorem,
which is proved in Appendix~\ref{AppDBimpliesStat}.
\begin{theorem} 
\label{DBimpliesStat}
If a \hyp{GDS}  $\bar{\Lambda}_t$ (Heisenberg picture) satisfies a $\hrhobgth$-\hyp{DBC}, 
then $\hrhobgth$ is invariant under the \hyp{GDS} $\Lambda_t$ (Schrödinger picture), 
\ie $\Lambda_t[\hrhobgth]= \hrhobgth$ or $\opL[\hrhobgth]=0$, equivalently.
\end{theorem}
In  the theorem above, the statement $\opL[\hrhobgth]=0$ follows from $d\Lambda_t[\hrhobgth]/dt=\opL[\Lambda_t[\hrhobgth]]$.
So, if a \hyp{GDS} satisfies a $\hrhobgth$-{\hyp{DBC}}, 
the quantum state $\hrhog^{\mathrm{th}}=\hrhobgth/\Tr(\hrhobgth)$ is a stationary state of the evolution. 
Therefore, in order to attain thermal equilibrium, 
it is enough that the superoperator $\bar{\Lambda}_t$ of a \hyp{GDS} satisfies a $\hrhobgth$-{\hyp{DBC}} 
for (the unnormalized Gibbs state) $\hrhobgth$ defined in~\eqref{GibbsstateH} 
with $\hH$ in~\eqref{Hform}. 

The equilibrium properties of a \hyp{GDS} can be extracted from
its relation with the so called modular automorphism group \cite{Carlen2017}:
\beq\label{automorphism}
\Xi_t[\hat A]=e^{\frac{\imath}{\hbar} \hH t}\,\hat A\,e^{-\frac{\imath}{\hbar} \hH t},
\eeq
with $\hat H$ in \eqref{Hform} and $t\in \Cset$. 
Of particular relevance will be the elements of the group given by the superoperator 
$ \Xi_{-\imath\hbar \beta}[\cdot]=(\hrhobgth)^{-1}\,\cdot\,\hrhobgth$, 
which existence is guaranteed for any finite value of $\beta$.
The relation between the $\hrhobgth$-{\hyp{DBC}} for an \hyp{GDS} 
and the above defined modular group is established in the following theorem:
\begin{theorem}
\label{GDScommod}
If a \hyp{GDS} satisfies a $\hrhobgth$-\hyp{DBC},
then $\bar{\Lambda}_t=e^{t\bar{\opL}}$ and $\bar{\opL}$ both commute with $ \Xi_t$
for all values of $t\in\Cset$.
\end{theorem}
This theorem was proved in \cite{Carlen2017} for \hyp{QDS} in finite-dimensional unital C$^*$-algebras 
and our demonstration for \hyp{GDS}s follows almost the same lines, see Appendix~\ref{AppGDScommod}.
\par
Now, due to the commutation relation between the Hamiltonians \eqref{Hform} and \eqref{Hformp},  
$[\hat H,\hat H_{\mathrm{eff}}] = 0$, we have that $\bar{\opL}_{\mathrm{U}}$ commutes with $\Xi_t$.
Therefore, the commutation of $\bar{\opL}=\bar{\opL}_{\mathrm{U}}+\bar{\opL}_{\mathrm{NU}}^{\mathrm G}$ 
with the automorphism $\Xi_t$ in (\ref{automorphism}) is equivalent to the following statement: 
$\bar{\opL}_{\mathrm{NU}}^{\mathrm G}$ commutes with $\Xi_t$.
The generator $\bar{\opL}_{\mathrm{NU}}^{\rm G}$ in~\eqref{linearLMEadjoint} for a \hyp{GDS} 
is an explicit function of both diffusion and dissipation matrices, $\DD$ and $\CC$ respectively.
The properties of theses matrices that stems from the fact that $\bar{\opL}_{\mathrm{NU}}^{\mathrm G}$ commutes with $\Xi_t$ 
is settled by the following theorem: 

\begin{theorem}
\label{TeocondDandC}
A \hyp{GDS} satisfies a $\hrhobgth$-{\hyp{DBC}} if and only if the diffusion and dissipation matrices, 
defined in Eqs.\eqref{defMats}, are such that 
\bea\label{conginvDDandCC}
\DD = \SST_t\DD\SST_t^\tp \,\,\, {\text{and}} \,\,\, \CC = \SST_t\CC\SST_t^\tp,
\eea
for $\SST_t$ ($t\in \Rset$) in Eq.\eqref{SSStdef}. 
\end{theorem}
This means that both matrices are invariant under a congruence relation through 
the symplectic matrix $\SST_t$. 


We begin the proof first noting that, for real values of $t$, 
the operator $e^{\frac{\imath}{\hbar} \hH t}$ in \eqref{automorphism}, 
with $\hH$ in~\eqref{Hform}, belongs to the metaplectic group {\rm Mp}($2n$,$\Rset$) of unitary operators 
and, consequently, is associated with the symplectic matrix 
$\SST_t^{-1} = \SST_{-t}$ defined by (\ref{SSStdef}) \cite{Ozorio1998,deGosson2006}.
%
So, in the Heisenberg picture, the action of these operators on the vector \eqref{opvecx} is described by
\bea \label{Ximent}
\Xi_{-t}[\hbx] = \SST_t^{-1}\hbx. 
\eea
Note that the above equation is equivalent to $\Xi_t[\hbx] = \SST_t\hbx$.
Using these actions, in Appendix~\ref{Appinvaderiva}, we prove that 
\beq
\label{invaderiva}
\Xi_{-t}\left[\pdv{}{\hbx^\tp}[\cdot]\right]=\pdv{}{\hbx^\tp}[\Xi_{-t}[\cdot]]\,\SST_t, 
\eeq
where $\Xi_{-t}=\Xi_t^{-1}$. This relation can be equivalently rewritten as 
$\Xi_{-t}[\partial /\partial \hbx[\cdot]]=\SST_t^\tp\,\partial /\partial \hbx[\Xi_{-t}[\cdot]]$.
%
Now, using Eq.\eqref{invaderiva}, we get 
\bse
\label{eqs}
\bea
\label{1steq}
\Xi_{-t}\left[\JJJ\hbx\pdv{}{\hbx}\, [\Xi_{t}[\cdot]] \right]&=&\SST_t^\tp\,\JJJ\hbx\pdv{}{\hbx}[\cdot]\,\SST_t,\\
\Xi_{-t}\left[\pdv{}{\hbx}\pdv{}{\hbx^\tp}[\Xi_t[\cdot]]\right]&=&\SST_t^\tp\pdv{}{\hbx}\pdv{}{\hbx^\tp}[\cdot]\SST_t,
\label{2ndeq}
\eea
\ese
where in~\eqref{1steq} we used the symplectic condition $\SST_t^{-1}=-\JJJ\SST_t^\tp\JJJ$.
Finally, inserting Eqs.\eqref{eqs} in \eqref{linearLMEadjoint} we attain
\bea
&&\Xi_{-t}[\bar{\opL}_{\mathrm{NU}}^{\rm G}[\Xi_t[\cdot]]]=\nonumber\\
&&=\tr\!\left(
\SST_t\CC\SST_t^\tp\JJJ\hbx\pdv{}{\hbx^\tp}\, [\hat O_t] 
+\frac{1}{2}\SST_t\DD\SST_t^\tp\pdv{}{\hbx}\pdv{}{\hbx^\tp} \hat O_t \right),
\eea
which is equal to $\bar{\opL}_{\mathrm{NU}}^{\rm G}[\cdot]$ in (\ref{linearLMEadjoint}) 
iff $\DD$ and $\CC$ satisfy Eqs.\eqref{conginvDDandCC}.
In summary, all these prove that the Eqs.\eqref{conginvDDandCC} are equivalent to the statement that $\bar{\opL}_{\mathrm{NU}}^{\rm G}$ 
commutes with $\Xi_t$ for any real value of $t$. 
However, due to $\Xi_{t^*}=\left(\Xi_t\right)^{{t^*}/{t}}$, 
$\bar{\opL}^{\rm G}_{\mathrm{NU}}$ must also commute with  $\Xi_{t^*}$ for  any complex value $t^*$.
We finish the demonstration of Theorem~\ref{TeocondDandC} noting that 
the commutation between $\bar{\opL}_{\mathrm{NU}}^{\rm G}$ and $\Xi_t$ with $t\in \Cset$ 
is tantamount to say that a \hyp{GDS} verifies a $\hrhobgth$-{\hyp{DBC}}, according to Theorem~\ref{GDScommod}.


As a consequence of the results in Theorem~\ref{TeocondDandC}, the Lindblad operators in~\eqref{LinearLindbOp} are restricted 
to a particular structure, since the congruence relations in \eqref{conginvDDandCC} are extended to the decoherence matrix $\GM$ in~\eqref{defGM}, 
due to the definitions in \eqref{defMats}, that is, $\GM=\SST_t\GM\SST_t^\tp$ for $\SST_t$ ($t\in \Rset$) 
in Eq.\eqref{SSStdef}.
%
Explicitly, the continuous-variable version of Theorem 3 from Alicki's work \cite{Alicki1976}, 
which deals with \hyp{QDS} in discrete Hilbert spaces, is a mere reformulation of our Theorem~\ref{TeocondDandC}:
\begin{theorem}
\label{ThoeremGMLindb}
A \hyp{GDS} satisfies a $\hrhobgth$-\hyp{DBC} 
iff the Lindblad operators describing the \hyp{GDS} are eigenoperators of the automorphism group $\Xi_{-t}$, \ie
\bse
\label{XimentLjLjplusn}
\bea
\Xi_{-t}[\hL_j]&=&e^{\imath\omega_j t}\hL_j,\\
\Xi_{-t}[\hL_{n+j}]&=&e^{-\imath\omega_j t}\hL_{n+j}=e^{-\imath\omega_j t}e^{-\frac{1}{2}\hbar\beta\omega_j}\hL_j^\dagger,
\eea
\ese
with $j=1,\ldots,n$ and $\omega_j>0$ are the eigenfrequencies (symplectic eigenvalues) 
of the Hessian matrix $\BB$ which defines, through the Hamiltonian Eq.\eqref{Hform}, 
the thermal equilibrium state $\hrhog^{\mathrm{th}}$ of the \hyp{GDS}.%
\end{theorem}%
The proof for this theorem stands on Theorem~\ref{TeocondDandC} and on Eq.(\ref{Ximent}), 
and some technical details are placed in Appendix~\ref{AppcoolformGM}.
In this appendix we prove that Eqs.\eqref{conginvDDandCC} are equivalent to write 
the decoherence matrix $\GM$ in the following characteristic form: 
\bea
\GM&=&\sum_{j=1}^{n} \left(|s_j|^2\lb_j\lb_j^\dagger+|r_j|^2 \lb^*_j(\lb^*_j)^\dagger\right)\nonumber\\
&=&(\QQ\SSSth)^{-1}(\ss\oplus\rr)(\ss^*\oplus\rr^*)(\QQ\SSSth)^{-\dagger},
\label{coolformGM}
\eea
where $\{\lb_j\}_{j=1,\ldots,n}$ are the eigenvectors of $\SST_t$, \ie
\beq
\label{Steigen}
\SST_t\,\lb_j=e^{\imath \omega_j t}\lb_j \quad (j=1,\ldots,n), \\
\eeq
the matrix $\rr := \text{diag}(r_1,\ldots,r_n)$ is the diagonal matrix satisfying  
\beq
\label{matbb}
|\rr|^2=\text{diag}(|r_1|^2,\ldots,|r_n|^2)=e^{-\frac{1}{2}\hbar\beta\ww}\,|\ss|^2,
\eeq
see Eq.\eqref{rr2sobress2} in Appendix~\ref{AppcoolformGM}; 
the diagonal matrix $\ww$ is defined in \eqref{diagsymplecBB} and 
contains the symplectic spectrum of the Hessian matrix $\BB$ of the Hamiltonian in~\eqref{Hform}; 
the matrix $|\ss|^2=\mbox{diag}(|s_1|^2,\ldots,|s_n|^2)$ is a real diagonal matrix 
with a particular temperature dependence. 
Although Theorem~\ref{ThoeremGMLindb} implies this dependence for any \hyp{GDS} satisfying 
its conditions, we will keep our track on the proof 
postponing the analysis of $|\ss|^2$ to Sec.\ref{ExamplesProperties}, 
see Eq.(\ref{si2value}).  
\par
Comparing the canonical form \eqref{coolformGM} with \eqref{defGM}, 
the Lindblad operators $\hL_k=\l_k^\tp\JJJ\hbx$ in (\ref{LinearLindbOp}), with $k=1,\ldots,K=2n$, 
correspond to the vectors
\bea 
\label{canonicalveclk}
\l_j=s_j\lb_j\, ,\quad \l_{n+j} &=& r_j\lb^\ast_j \quad(j=1,\ldots,n).
\eea
From \eqref{Steigen}, 
$\SST_t\l_{n+j}=\SST_t\lb_j^\ast=e^{-\imath \omega_j t} r_j \lb_j^\ast$, consequently 
$\Xi_{-t}[\hL_j]= 
\l_j^\tp\SST_t^\tp\JJJ\hbx= s_j\lb_j^\tp\SST_t^\tp\JJJ\hbx=e^{\imath\omega_j t}s_j\lb_j^\tp\JJJ\hbx=e^{\imath\omega_j t}\hL_j$ 
and 
$\Xi_{-t}[\hL_{n+j}]=
r_j\lb^\dag_{j} \SST_t^\tp\JJJ\hbx=e^{\imath\omega_j t}\,r_j\,\lb^\dag_{j}\JJJ\hbx=e^{-\imath\omega_j t}e^{-\frac{1}{2}\hbar\beta\omega_j}\hL^\dagger_{j}$,
where we used~\eqref{Ximent} and \eqref{matbb} and the symplectic condition 
$\JJJ\SST_t^{-1}=\SST_t^\tp\JJJ$. 
With all these, we finish the proof of Theorem~\ref{ThoeremGMLindb}.
\par
Here, two important observations are in order. 
First, the decoherence matrix \eqref{coolformGM} could be written as $\GM=\Sig\Sig^\dagger$,  
where 
\beq
\label{defSig}
\Sig=(\QQ\SSSth)^{-1}(\ss\oplus \ss e^{-\frac{1}{2}\hbar\beta\ww})\WW
\eeq 
and $\WW\in \Cset^{2n\times 2n}$  is
an arbitrary unitary matrix. 
Therefore, alternatively we can use the columns vectors $\l^\prime_j$ of the matrix $\Sig$ 
to define new Lindblad operators $\hL_k^\prime=(\l^\prime_k)^\tp\JJJ\hbx$ with $k=1,\ldots,2n$. 
Notwithstanding, it is straightforward to check that transformation \eqref{defSig} corresponds to
\begin{equation}
\label{eq:unitarytransformation}
\hL_k \longrightarrow \hL^\prime_k = \sum^{2n}_{j=1} \WW_{kj} \,  \hL_j  
\end{equation} 
and the arbitrariness introduced by $\WW$ in~\eqref{defSig} is equivalent to a well known symmetry of 
any \hyp{QDS} (see \eg \cite[Sec.3.2.2]{Breuer2002}): 
the \hyp{LME} \eqref{eq:mastereq} is invariant under the unitary transformation in~\eqref{eq:unitarytransformation} 
of the Lindblad operators. Therefore, this symmetry also holds for a \hyp{GDS}. Thus, the semigroup dynamic
 associated to the new set of Lindblad operators, $\hL_k^\prime$, is exactly the same as the one generated by the old set, \ie
$\hL_k$ in (\ref{LinearLindbOp}) with $\lb_k$ in \eqref{canonicalveclk}. 
Secondly, Theorem~\ref{ThoeremGMLindb} shows that it is enough to consider only $n$ Lindblad operators 
to describe an \hyp{GDS} which satisfies a $\hrhobgth$-\hyp{DBC}, 
although we are dealing with infinite dimensional quantum systems.  
In this sense, \hyp{GDS} are like \hyp{QDS} in finite-dimensional systems.
\par
Properties of the environment,
characterized by the matrices $\DD$ and $\CC$, can be extracted 
from the canonical form of the decoherence matrix $\GM$ in~\eqref{coolformGM}.
To this end, we conveniently rewrite $\GM$ in~\eqref{coolformGM} as
\beq
\label{rewriteGM}
\GM=\frac{\DD}{\hbar}+\imath \CC=(\SSSth)^{-1}\left(\Lambt_r-\imath \JJJ\Lambt_i\right)(\SSSth)^{-\tp},
\eeq
where we define the diagonal positive defined matrices
\bse
\label{FormLambt}
\bea
\Lambt_r&=&  \frac{|\ss|^2+|\rr|^2}{2} \oplus  \frac{|\ss|^2+|\rr|^2}{2}\nonumber\\
&=&\frac{|\ss|^2}{2}(\In+e^{-\hbar\beta\ww})\oplus\frac{|\ss|^2}{2}(\In+e^{-\hbar\beta\ww}),
   \label{defDd}\\
\text{and}\;\, \Lambt_i&=&   \frac{|\ss|^2-|\rr|^2}{2} \oplus  \frac{|\ss|^2-|\rr|^2}{2}\nonumber\\
&=&\frac{|\ss|^2}{2}(\In-e^{-\hbar\beta\ww})\oplus\frac{|\ss|^2}{2}(\In-e^{-\hbar\beta\ww}),
     \label{defJCd}
\eea
\ese
using the notation $|\ss|^2=\text{diag}(|s_1|^2,\ldots,|s_n|^2)$,  
$|\rr|^2=\text{diag}(|r_1|^2,\ldots,|r_n|^2)$
and the matrix relation in \eqref{matbb}.
Therefore, the diffusion and dissipation matrices are
\bse
\label{DDandCC-DBC}
\bea
\DD&=&(\SSSth)^{-1}\hbar\Lambt_r(\SSSth)^{-\tp},
\label{DD-DBC}\\
\CC&=&(\SSSth)^{-1} \JJJ^\tp\Lambt_i(\SSSth)^{-\tp}.
\label{CC-DBC}
\eea
\ese
Note that each pair of matrices $\DD$ and $\CC$ uniquely determine
a \hyp{GDS}. Since these only depend on the real matrix $|\ss|^2$, 
it is enough to choose  a real matrix $|\ss|=\text{diag}(|s_1|,\ldots,|s_n|)$ 
instead of   a complex matrix $\ss$ in~\eqref{coolformGM} with $\rr$ in~\eqref{matbb}. 
In this way the relations in~\eqref{canonicalveclk} change to 
\bea 
\label{canonicalveclk2}
\l_j=|s_j|\lb_j, \,\, \l_{n+j} = |s_j|\,e^{-\frac{1}{2}\hbar\beta\omega_j}\lb^\ast_j \,\,\,\, (j=1,\ldots,n), 
\eea
which gives the expressions for the coefficients 
of the Lindblad operators in~\eqref{LinearLindbOp}.
\par 
The structure of $\DD$ and $\CC$ in~\eqref{DDandCC-DBC} is determined by 
the matrices $\SSSth$, $\ww$, and $|\ss|$.
It is worth to note that, according to Eq.\eqref{SpecJB},  
the matrices $\SSSth$ and $\ww$ can be extracted 
from the ordinary diagonalization of the Hamiltonian matrix $\JJJ\BB$, where $\BB$
is the Hessian matrix of the Hamiltonian $\hH$ in~\eqref{Hform}, that 
defines the thermal equilibrium state $\hrhog^{\mathrm{th}}$ in \eqref{GibbsstateH}. 
However, we will see in Sec.\ref{ExamplesProperties} that $|\ss|$ depends on the coupling constants of the system and the environment.
Thus, each matrix $|\ss|$, corresponding to different coupling constants, defines one different pair of diffusion and dissipation matrices through Eqs.\eqref{DDandCC-DBC}.
The dynamics of the \hyp{GDS}s associated to these matrices is different because 
each one corresponds to different Lindblad operators,
which are not associated with the symmetry in~\eqref{eq:unitarytransformation}.
Nonetheless, all these \hyp{GDS}s have the same 
thermal equilibrium state $\hrhog^{\mathrm{th}}$. This is checked in Appendix~\ref{AppcoolformGM} through 
the symplectic diagonalization of $\VVth$ in~\eqref{diagsympVVth}, where the matrix $\frac{1}{2}\Lambt_r\Lambt_i^{-1}=\kkth\oplus\kkth$ is the direct sum of the symplectic spectrum of $\VVth$ that does not depend on $|\ss|$.

\section{The master equation of a GDS satisfying a 
$\hrhobgth$-{\rm DBC}. }
\label{secQOME}
Here, we prove that the master equation of a \hyp{GDS} that satisfies a $\hrhobgth$-\hyp{DBC}, 
\ie satisfying all theorems in last section, has the form of the Quantum Optical Master Equation (\hyp{QOME}), 
see Eq.(5.4.14) in \cite{GardinerZoller2000}:
\bea
\dv{\hrho_t}{t} &=& -\frac{\imath}{  \hbar}[\hHo, \hrho_t ] \nonumber \\
&&+
\sum_{k} \frac{\bar{\gamma}_k}{2}(\bar{n}_k+1) 
\left( 2 \hXme_k \hrho_t \hXma_k - \{\hXma_k \hXme_k, \hrho_t\}\right)\nonumber\\
&&+
\sum_{k} \frac{\bar{\gamma}_k}{2}\bar{n}_k 
\left( 2 \hXma_k \hrho_t \hXme_k - \{\hXme_k \hXma_k, \hrho_t\} \right),
\label{eqQOME}
\eea
where $\bar{\gamma}_k$ are the coupling constants between the bath and the system, 
$\bar{n}_k=(e^{\hbar\beta\tilde{\omega}_k}-1)^{-1}$
is the Planck factor with $\tilde{\omega}_k>0$, 
and the operators $\hX^{\pm}_k$ are eigenoperators of $\hHo$, \ie
\beq
\label{commhXmemaQOME}
[\hHo,\hX^{\pm}_k]=\pm\hbar\tilde{\omega}_k \hX^{\pm}_k,
\eeq
so $(\hXme_k)^\dagger=\hXma_k$. 
\par
In order to rewrite the master equation of a \hyp{GDS} satisfying a $\hrhobgth$-\hyp{DBC}, 
first we rewrite the Lindblad operators in (\ref{LinearLindbOp}) as 
\bse
\label{LindbopGDSwDBC}
\bea
\hL_j&=&|s_j|\hXme, \\
\hL_{n+j} &=&e^{-\frac{1}{2}\hbar\beta\omega_j}\hL^\dagger_j=|s_j|e^{-\frac{1}{2}\hbar\beta\omega_j} \hXma_j
\eea
\ese
with 
\beq
\label{defhXmahXme}
\hXme_j=\lb_j^\tp\JJJ\hbx\quad\text{and}\quad \hXma_j=(\hXme)^\dagger=(\lb^\ast_j) ^\tp\JJJ\hbx, 
\eeq
when employing Eq.\eqref{canonicalveclk2} for $j = 1,...,n$. 
Therefore, using these expressions for the Lindblad operators, 
the master equation in (\ref{eq:mastereq}) becomes
\bse
\label{MEGDSwDBC}
\bea
\!\!\!\!\dv{\hrho_t}{t} &=& -\frac{\imath}{  \hbar}[\hHp, \hrho_t ]\nonumber\\
+&& \!\!\!\!\!\!\!\!
\sum_{j=1}^n \frac{|s_j|^2}{2\hbar} 
( 2 \hXme_j \hrho_t \hXma_j - \{\hXma_j \hXme_j, \hrho_t\})
\label{MEGDSwDBCa}\\
+&&\!\!\!\!\!\!\!\!
\sum_{j=1}^n \frac{|s_j|^2}{2\hbar}e^{-\hbar\beta\omega_j}\!
( 2 \hXma_j \hrho_t \hXme_j - \{\hXme_j \hXma_j, \hrho_t\}).
\label{MEGDSwDBCb}
\eea
\ese 
In Appendix~\ref{AppLindQDBC}, we prove that $\hX_j^{\pm}$ are eigenoperators of $\hH$
in~\eqref{Hform}, \ie
\beq
\label{commhXmema}
[\hH,\hX_j^{\pm}]=\pm\hbar\omega_j\hX_{j}^\pm \,\,\,\,\,\,  (j=1,\ldots, n),
\eeq
while in Appendix~\ref{Appmatss} there is a proof for  
\beq
\label{si2value}
|s_j|^2=\hbar\bar{\gamma}_j\,\bar{n}_j\,e^{\hbar\beta\omega_j}=\hbar\bar{\gamma}_j\,(\bar{n}_j+1),
\eeq
where now the Planck factor is
\beq
\label{Planckfactor}
\bar{n}_k=(e^{\hbar\beta\omega_k}-1)^{-1}.
\eeq
With all these, we show that the master equation~\eqref{MEGDSwDBC}  has the structure of 
a \hyp{QOME} like the one in \eqref{eqQOME}. 
\par
Using Eq.\eqref{diagsymplecBB} and writing $\hat U_{\SSSth}$ for the metaplectic operator 
associated with the symplectic matrix $\SSSth$, 
\ie $\hat U_{\SSSth}^\dagger\hbx\hat U_{\SSSth}= \SSSth\hbx$, one can find
\beq
\label{hHhHho}
\hH=\frac{1}{2}\hbx^\tp\BB\hbx=\hat U_{\SSSth}^\dagger \hH_{{\rm ho}}\hat U_{(\SSSth)}, 
\eeq 
where 
\[
\hH_{{\rm ho}} :=  \frac{1}{2}\hbx^\tp (\ww\oplus\ww) \hbx = 
\sum_{j = 1}^{n} \frac{\omega_j}{2} (\hat p_j^2 + \hat q_{j}^2)
\]
is the Hamiltonian of a multimode harmonic oscillator. 
The eigenequation for $\hH_{\rm ho}$ is
\beq
\label{Enspectrum}
\hH_{\rm ho}\ket{\bf n} = E_{\bf n}\ket{\bf n}, \,\,\,  E_{\bf n}=\sum_{l=1}^n\hbar\omega_l\left(n_l+\frac{1}{2}\right),
\eeq
where 
$\ket{\bf n}=\ket{n_1}\otimes\ldots\otimes\ket{n_n}$, ${\bf n}=(n_1,\ldots,n_n)$
with $n_l=0,\ldots,+\infty$. 
Therefore, due to the similarity relation between $\hH_{{\rm ho}}$ and $\hH$, 
the eigenequation of $\hH$ is
$\hH\ket{\phi_{\bf n}}=E_{\bf n}\ket{\phi_{\bf n}}$,
with
\beq
\label{eigenvhH}
\ket{\phi_{\bf n}}=\hat U_{\SSSth}^\dagger\ket{{\bf n}}.
\eeq
\par 
Now, from~\eqref{commhXmema}, it is straightforward to verify that $\hXma_j$ and $\hXme_j$ are the 
creation and annihilation operators of a quantum $\hbar \omega_j$ associated to the states $\ket{\phi_{\bf n}}$; 
that is, $\hX^{\pm}_j \ket{\phi_{\bf n}}=\alpha_{{\bf n}_j^\pm}|\phi_{{\bf n}_j^\pm}\rangle$ 
is an eigenvector of $\hH$ with energy $E_{{\bf n}_j^\pm}$ such that ${\bf n}_j^\pm=(n_1,\ldots,n_j\pm1,\ldots,n_n)$. 
%
Since $\hX_j^\pm$ satisfies the commutation relation~\eqref{commhXmema}, 
the operator $\hXma_j\hXme_j$ is the number operator in 
the eigenbasis $\{\ket{\phi_{\bf n}}\}$, \ie $\hXma_j\hXme_j\ket{\phi_{\bf n}}=n_j\ket{\phi_{\bf n}}$, 
therefore $\bar{\alpha}_{{\bf n}_j^+}=\sqrt{n_j+1}$ and $\bar{\alpha}_{{\bf n}_j^-}=\sqrt{n_j}$
\footnote{This can be checked through a lengthy, but not difficult, 
calculation using the definition of $\hX_j^\pm$ in~\eqref{defhXmahXme} and $\ket{\phi_{\bf n}}$ in~\eqref{eigenvhH}.}.
\par
It is worth to note that we consider the same mode structure for $\hH_{\rm ho}$ and $\hH$, 
what changes is the nature of the stationary states, \ie while $\ket{\bf n}$ are separable states, 
$\ket{\phi_{\bf n}}$ could be entangled with respect to the considered mode structure.
In an analogous way, we can rewrite the thermal equilibrium state in~\eqref{GibbsstateH} 
as
\bea
\hrhog^{\mathrm{th}}&=&\frac{1}{{\cal Z}^{\mathrm{th}}}e^{-\frac{\beta}{2}\hbx^\tp (\SSSth)^{\tp}(\ww\oplus\ww) \SSSth \hbx }\nonumber\\
&=&\hat U_{\SSSth}^\dagger \hrhog^{\mathrm{th}}_{\rm ho} \hat U_{\SSSth},
\label{hrhogdeco}
\eea
where we recognize in 
\beq
\hrhog^{\mathrm{th}}_{\rm ho}=\frac{1}{{\cal Z}^{\mathrm{th}}}e^{-\frac{\beta}{2}\hbx^\tp (\ww\oplus\ww) \hbx }
\eeq
the Gibbs's state of the multimode harmonic oscillator with Hamiltonian $\hH_{\rm ho}$ in \eqref{hHhHho}.
The state $\hrhog^{\mathrm{th}}_{\rm ho}$ is manifestly separable, 
so the possible entanglement of the state $\hrhog^{\mathrm{th}}$ 
is due to the action of the unitary operation $\hat U_{(\SSSth)}$,
where $\SSSth$ is the matrix of the symplectic diagonalization of the Hessian matrix $\BB$, as shown in Eq.(\ref{diagsymplecBB}).
\par
Noteworthy that usually in the derivation of the \hyp{QOME},
it is assumed that the commutation relation~\eqref{commhXmemaQOME} is valid \cite{GardinerZoller2000}. 
Here, the quantum detailed balance condition for a \hyp{GDS} 
shows that commutation relation~\eqref{commhXmema} must be valid instead.
%
When $\SSSth$ is the identity and $\hHp=\hH=\hH_{\rm ho}$,
we have that $\hXme_j=\hat a_j$ and $\hXma_j=\hat a_j^\dagger$ 
are the usual creation and annihilation operators of the multimode quantum harmonic oscillator. 
Then, in this case the \hyp{QOME}~\eqref{MEGDSwDBC} is the well known master equation of a multimode damped 
harmonic oscillator~\cite{GardinerZoller2004}.  \\

\section{Additional properties of GDS\lowercase{s} satisfying a $\hrhobgth$-{\bf DBC}}
\label{ExamplesProperties} 

\subsection{The temperature dependence of \texorpdfstring{$\DD$}{D} and \texorpdfstring{$\CC$}{C}}
\label{tempDDandCC}

The final form for the diffusion and dissipation matrices of GDSs that satisfy a $\hrhobgth$-\hyp{DBC} 
is obtained replacing~\eqref{si2value} into~\eqref{DDandCC-DBC}, so 
\bse
\label{DDandCCdef}
\bea
\DD&=&\hbar^2(\SSSth)^{-1} 
(\gaga \kkth\oplus \gaga \kkth)\;(\SSSth)^{-\tp}, \label{DDdefinitiva} \\
\CC&=&\frac{\hbar}{2} (\SSSth)^{-1}\;\JJJ^\tp(\gaga\oplus\gaga) (\SSSth)^{-\tp},
\label{CCdefinitiva}
\eea
\ese
where $\kkth$ is given in~\eqref{VVthsymdiag2} and $\gaga$ is the diagonal matrix containing 
the coupling constants, \ie
\beq \label{eqcoupmat}
\gaga=\text{diag}(\bar{\gamma}_1,\ldots,\bar{\gamma}_n).
\eeq%
It is worth to remember here that $\SSSth$ and $\kkth$ are ultimately determined by $\BB$, the Hessian matrix of 
$\hH$ that defines the thermal equilibrium state. 
However, the coupling constants can take different values, 
thus defining different pair of matrices, $\DD$ and $\CC$, 
all having the same equilibrium state. These are the multiples solutions of Eq.\eqref{Vthposta}.
\par
Note that here we consider that the coupling constants do not depend on temperature.
This is consistent with the fact that the Fokker-Planck equation \eqref{FPEGDS} for Gaussian states corresponds to 
an Ornstein--Uhlenbeck  process \cite{Risken1996}, 
where the drift term $\pdv{\JJJ\CC\x}{\x^\tp}=\tr(\JJJ \CC)$ corresponds to a drift force $-\JJJ\CC\x$ 
which does not depend on the temperature \cite{Toscano2021}.
From Eq.\eqref{Comsetdiag2}, we can write $\DD=\hbar (\SSSth)^{-1} (\dd\oplus\dd)(\SSSth)^{-\tp}$, where we use
that $\SSSth$ is symplectic and that $\QQ^\dagger \,(\imath \dd\oplus (-\imath \dd ))\,\QQ=\hbar\JJJ(\dd\oplus\dd)$.
Analogously, from Eq.\eqref{Comsetdiag3}, we can write $\CC=(\SSSth)^{-1}\JJJ^\tp (\jc\oplus\jc)(\SSSth)^{-\tp}$,
where $\QQ^\dagger (\jc\oplus  \jc) \QQ=\jc\oplus\jc$. Comparing with Eqs.\eqref{DDandCCdef}, 
we have  $\dd=\hbar\gaga\,\kkth$ and $\jc=\frac{\hbar}{2}\gaga$, so $\dd=2\jc\,\kkth$. 
Therefore, the diffusion and dissipation matrices of a \hyp{GDS} satisfying a \hyp{QDBC} verify the 
condition \eqref{cucamuruca}; this condition guarantees that the \hyp{GDS} has a thermal equilibrium state. 
\par 
For the validity of the reciprocal implication 
(every \hyp{GDS} that attains thermal equilibrium must satisfy a \hyp{QDBC}), 
we need to allow an arbitrary temperature dependence for the coupling constants in the matrix $\gaga$. 
This arbitrariness is clear taking into account \eqref{cucamuruca}, 
where $\dd=2\jc\,\kkth$; now $\jc$ can depend on temperature 
still matching \eqref{CCdefinitiva} with $\jc=\frac{\hbar}{2}\gaga$ 
for an arbitrary dependence of $\gaga$ on temperature.
\\
\par
\subsection{The high and low temperature limits}
\label{handltemplimit}
The high temperature limit is obtained when considering 
$\hbar \beta \|\ww\|\ll 1$ in~\eqref{DDdefinitiva}, where $\|\ww\|={\max}\{\omega_1,...,\omega_n\}$,
so one can write $\kkth\approx (\hbar\beta\ww)^{-1}+\frac{1}{12}\hbar\beta\ww+{\cal O}(\hbar \beta \ww )^3$ 
and neglecting higher order terms, we write 
\beq
\DD\approx\hbar\beta^{-1}(\SSSth)^{-1} (\gaga \ww^{-1}\oplus \gaga\ww^{-1} )(\SSSth)^{-\tp}.  
\eeq
Inserting this into Eq.(\ref{Vthposta}) with $\CC$ in (\ref{CCdefinitiva}),  
using the symplectic condition for $\SSSth$, and the symplectic diagonalization in (\ref{diagsymplecBB}), 
one obtains
\beq
\VV \approx ({\hbar \beta}\BB)^{-1}. 
\eeq
This limit is also called classical limit and the Wigner function in (\ref{eq:WignerGauss}) 
for this covariance matrix is the classical Boltzmann factor 
$\exp[-\beta H]$ for the classical version (Wigner symbol) 
of Hamiltonian (\ref{Hform}) \cite{Nicacio2021Weyl}. 
\par
The low temperature limit corresponds to  $\hbar \beta \|\ww\|\gg 1$ and in this case 
$\kkth=\frac{1}{2}\In+{\cal O}(e^{-\hbar \beta \ww })$. From~\eqref{DDandCCdef},
\beq
\label{DDlowtemp}
\DD \approx \hbar \frac{1}{2} (\SSSth)^{-1}(\SSSth)^{-\tp}\JJJ\CC.
\eeq
Consequently, 
\beq
\VVth\approx\frac{1}{2} (\SSSth)^{-1}(\SSSth)^{-\tp},
\eeq
and the thermal equilibrium state corresponds to a pure Gaussian state, 
$\hrhog^{\mathrm{th}}\approx\dyad{\phi_{{\bf n}={\bf 0}}}{\phi_{{\bf n}={\bf 0}}}$, 
where $\ket{\phi_{{\bf n}={\bf 0}}}$ is the ground state of the Hamiltonian $\hH$ in~\eqref{Hform}. \\
\par

\subsection{The pure diffusive regime}\label{diffusivereg} 

This regime corresponds to the limit 
$\bar{n}_j\rightarrow +\infty$ (\ie $\beta \rightarrow 0$) together with 
$\bar{\gamma}_j\rightarrow 0$, such that $\bar{n}_j\bar{\gamma}_j=\bar{c}_j$
are constant values for $j=1,\ldots,n$. 
In such limit, the coeficients in~\eqref{MEGDSwDBCa} become equal 
to the ones in~\eqref{MEGDSwDBCb} for the master equation of the \hyp{GDS}, see also Eq.(\ref{eqQOME}).
For the difussion and dissipation matrices, respectively, we have
\bse
\label{DandCdifreg}
\bea
\DD&=& \hbar^2(\SSSth)^{-1}\,\ccb\oplus\ccb\,(\SSSth)^{-\tp},\\
\CC&=& 0,
\eea
\ese
where $\ccb=\text{diag}(\bar{c}_1,\ldots,\bar{c}_n)$. 
In this case the \hyp{GDS} has no thermal equilibrium solution 
because $\DD$ and $\CC$ do not satisfy Eq.\eqref{Vthposta}. 
\par
It is worth to note that the master equation in~\eqref{MEGDSwDBC} 
has no pure dissipative regime, \ie $\DD=0$, since Eq.~\eqref{DDplusiCC} is violated. 
However, if $\hbar$ represents an effective Planck constant in the master equation 
of the \hyp{GDS} in~\eqref{MEGDSwDBC}, the semiclassical limit $\hbar \ll 1$ and 
the low temperature condition $\hbar \beta \|\ww\|\gg 1$ guarantee the validity of \eqref{DDlowtemp},  
thus the contributions from $\DD$ can be effectively neglected 
when compared to those coming from $\CC$; 
in this case the evolution is thus dominated by dissipation \cite{Toscano2005}.  \\

\section{The role of $\hat H_{\rm eff}$ in thermalization}
\label{examples}

So far, we completely answer the question of which kind of environments, 
characterized by the diffusion and dissipation matrices $\DD$ and $\CC$, 
allows the existence of a given thermal equilibrium state in a \hyp{GDS}. 
This state is given by a Gaussian Gibbs state, 
$\hrhog^{\mathrm{th}}= e^{-\beta\hH}/ {\cal Z}^{\mathrm{th}}$,
corresponding to a quadratic Hamiltonian $\hH$ of positive-elliptic type, which 
is completely characterized by its covariance matrix $\VVth$ such that $\JJJ\VVth$ 
belongs to the commuting set $\{\JJJ\VVth,\JJJ\DD,\JJJ\CC\}$. 
\par
However, it is interesting to analyze this result from the perspective of a fixed \hyp{GDS} 
that governs the system to thermal equilibrium.
In this case, the commuting matrices $\JJJ\DD$ and $\JJJ\CC$ determine $\JJJ\BB$ 
through the relation~\eqref{Vthposta} and~\eqref{VVthJfuncofJB}.
Then, it is clear that the effective action of the environment, through the non-unitary part of the evolution in~\eqref{linearLME}, 
is to confine the system in phase space (with coordinates $\x$) 
around the the equilibrium point $\expval{\hbx}^{\mathrm{th}}=0$.
The energetic balance between the environment and the system is described by the quantum positive-elliptic Hamiltonian $\hH$, 
which is dynamically stable (with spectrum necessarily discrete with bounded eigenstates). 
If we admit that every  \hyp{GDS} that leads to thermal equilibrium satisfies a \hyp{QDBC}, the form of the master equation 
in~\eqref{MEGDSwDBC} together with~\eqref{commhXmema} and~\eqref{si2value}, helps to understand in detail the
energetic balance of the thermal equilibrium process.  Indeed, this process is determined by the action of the eigenoperators $\hX_j^{\pm}$ of $\hH$, 
which promote transitions between the energy levels of this Hamiltonian as a consequence of the interaction with the environment. 
This contrasts with the usual derivation \cite{GardinerZoller2000} 
of the \hyp{QOME} in~\eqref{eqQOME}, 
where $\{\hX_j^{\pm}\}$ is considered as a basis of eigenoperators of $\hHo$, 
which expands the algebra of operators acting on the system \cite{GardinerZoller2000}.
\par
The induced environmental confinement process of the system in thermal equilibrium
is not influenced by the unitary dynamics generated by $\hHp$, 
since the commuting set of matrices $\{\JJJ\VVth,\JJJ\DD,\JJJ\CC\}$, 
which ultimately determines the thermal equilibrium through \eqref{Vthposta}, 
do not depend on the Hessian matrix $\BBp$ of $\hHp$.
However, there are some of these Hamiltonians that allows the process of thermalization to occur, 
these are determined by the condition $[\hH,\hHp]=0$. 
Using Eq.\eqref{commHam} and the results in Appendix~\ref{AppComRel}, 
all these Hamiltonians are written as  
\beq\label{aux12}
\hHp=\frac{1}{2}\hbx^\tp\BBp\hbx=\hat U_{\SSSth}^\dagger \hH^\prime_{{\rm ho}}\hat U_{(\SSSth)},
\eeq
where we employed the same reasoning performed in Eq.\eqref{hHhHho} 
with $\BBp=(\SSSth)^\tp(\xx\oplus\xx)\SSSth$. 
The Hamiltonian $\hH^\prime_{{\rm ho}}$ is the one of a multimode harmonic oscillator, \ie
\[
\hH^\prime_{{\rm ho}} :=  \frac{1}{2}\hbx^\tp (\xx\oplus\xx) \hbx = 
\sum_{j = 1}^{n} \frac{\lambda_j}{2}  (\hat p_j^2 + \hat q_{j}^2)
\]
with $\xx=\text{diag}(\lambda_1,\ldots,\lambda_n)$ and $\lambda_j \in \Rset$, $j = 1,..., n$.

Likewise $\hH$, the Hamiltonian $\hHp$ in (\ref{aux12}) is also elliptic, or dynamically stable. 
However, contrary to the multimode harmonic oscillator $\hH_{{\rm ho}}$ in the expression~\eqref{hHhHho}, 
the arbitrary frequencies $\lambda_j$ of $\hH^\prime_{{\rm ho}}$ can be positive, negative, 
or even null. 
Consequently, $\hHp$ can be the null Hamiltonian and the system still thermalizes 
to the same state $\hrhog^{\mathrm{th}}$. 
Note, however, that taking a null Hamiltonian is quite different of considering the interaction representation 
of the \hyp{QOME} in~\eqref{eqQOME}, where the density operator turns to  
$\hrho_t^I=e^{\frac{\imath}{\hbar}\hHo t}\hrho_t e^{-\frac{\imath}{\hbar}\hHo t}$, since 
the frequencies $\tilde{\omega}_k$ still correspond to $\hHo$ through the Planck factor $\bar{n}_k$. 
\par
 It is worth to note that the operators $\hX^{\pm}_k$ are eigenoperators of $\hH$ 
when the master equation~\eqref{MEGDSwDBC} of a \hyp{GDS} thermalizes, see Eq.~\eqref{commhXmema}, 
but they are also eigenoperators of $\hHp$, \ie 
$[\hH,\hX_j^{\pm}]=\pm\hbar\lambda_j\hX_{j}^\pm$ with $j=1,\ldots,n$.
Nevertheless, the frequencies $\omega_k$ in the master equation~\eqref{MEGDSwDBC} are associated with $\hH$. 
Thus, the energetic balance of the thermalization process is associated 
with transitions in the Hamiltonian $\hH$ associated with the confinement of the system.

\section{Conclusions}\label{conclusions}
When a \hyp{GDS} has a thermal equilibrium state $\hrhog^{\mathrm{th}}$, 
every initial condition of the evolution ends up on this state. 
Without lost of generality, we can choose $\expval{\hbx}^{\mathrm{th}}=0$, 
so the thermal state $\hrhog^{\mathrm{th}}$ must have the form in~\eqref{GibbsstateH} with 
$\hH$ in~\eqref{Hform}, and the Hamiltonian of the unitary part of the evolution for the GDS given by
$\hHp$, of the form in~\eqref{Hformp} such that $\hH=\hHp$ or $[\hH,\hHp]=0$. 
Another prerequisite for $\hrhog^{\mathrm{th}}$ came from the normalization of a density operator: 
in order to have unity trace, the Hessian matrix $\BB$ of $\hH$ must be positive definite \cite{Nicacio2021Weyl}. 
This means that $\hH$ is elliptic, \ie
its classical counterpart in~\eqref{quadclassham} generates a Hamiltonian flow 
given by the symplectic matrix $\SST_t $ in~\eqref{SSStdef} around the elliptic fixed point $\x=0$.
Therefore, $\hH$ is dynamically stable having a discrete energy spectrum corresponding to bound states.
\par
The non-unitary part of the evolution of a \hyp{GDS} is determined by the diffusion and dissipation matrices, 
$\DD$ and $\CC$, respectively. 
These are the real and imaginary parts, respectively, of the decoherence matrix $\GM$, see Eq.(\ref{defGM}).
In~\eqref{Vthposta}, we show the relation between the covariance matrix $\VVth$ of $\hrhog^{\mathrm{th}}$ 
and the matrices $\DD$ and $\CC$ for every \hyp{GDS} with a thermal equilibrium state, 
which has appeared before in \cite{Toscano2021} in a narrower context.
Further, we show that $\VVth$ is completely determined by the symplectic matrix $\SSSth$ in~\eqref{VVthsymdiag},  
which diagonalizes symplectically the Hessian matrix $\BB$ as shown in \eqref{diagsymplecBB}, 
and by the symplectic spectrum of $\BB$, which contains the eigenfrequencies $\omega_j>0$ 
of the system Hamiltonian. 
\par
However, Eq.\eqref{Vthposta} does not establish neither the temperature dependence of $\DD$ and $\CC$, nor 
characterize all the different GDSs that have the same thermal equilibrium state $\hrhog^{\mathrm{th}}$.
This deficiency can be fixed extending Alicki's \hyp{QDBC} (developed for systems described by discrete Hilbert spaces) 
to $n$-mode bosonic systems.
We have shown how to implement this extension and the result is that 
all GDSs verifying a \hyp{QDBC} have a thermal equilibrium state whose dynamics is
characterized by symplectically invariant diffusion and dissipation matrices,  
see Eq.(\ref{conginvDDandCC}). 
This condition allows the characterization of the Lindblad operators of a 
\hyp{GDS} satifying a \hyp{QDBC} in Theorem~\ref{ThoeremGMLindb}.
\par
The corresponding master equation 
has the structure of a Quantum Optical Master Equation (QOME),  
see Eq.~\eqref{MEGDSwDBC} together with Eqs.~\eqref{commhXmema} and~\eqref{si2value}.
As a consequence, the matrices $\DD$ and $\CC$ have the specific structure in \eqref{DDandCCdef}, 
which sets explicitly their dependence on temperature. 
These expressions also show that different GDSs with the same thermal equilibrium state 
differ from each other only on the value of the coupling constants, 
namely $\bar{\gamma}_j$ in~\eqref{si2value}, 
because $\SSSth$ is determined by the symplectic diagonalization of $\BB$, \ie the Hessian matrix of $\hH$.
Usually the coupling constants $\bar{\gamma}_j$ do not depend on temperature, however we show that
if we allow these to have an arbitrary temperature dependence, 
then it is possible to say that every \hyp{GDS} with thermal equilibrium satisfies a \hyp{QDBC}.
\par
We also have shown that it is possible to define a pure diffusive regime 
for any \hyp{GDS} satisfying the \hyp{QDBC},
where the \hyp{GDS} lost its stationary solution. 
This is a specific example of the necessity to balance diffusion and dissipation to achieve a
stationary solution in any quantum dynamical semigroup \cite{Toscano2021}.
\par
Finally, we also show that the contribution of $\hHp$ to the dynamics of a \hyp{GDS}  
has no influence in the thermalization process. However, by determining all $\hHp$ such that
$[\hH,\hHp]=0$, we have determined all Hamiltonians that allows thermalization. 
These Hamiltonians are also dynamically stable as $\hH$, 
however with arbitrary eigenenergy frequencies. 
Therefore, the energetic balance of the thermalization process, 
determined by $\hH$, is completely independent of $\hHp$. 
This marks a fundamental difference between the  \hyp{QOME} for a \hyp{GDS} satisfying 
a \hyp{QDBC} and the usual \hyp{QOME}~\cite{GardinerZoller2000}. \\

\begin{acknowledgments}
The authors are members of the Brazilian National Institute of Science 
and Technology for Quantum Information [CNPq INCT-IQ (465469/2014-0)].  
FN acknowledges partial financial support from the Brazilian agency CAPES 
[PrInt2019 (88887.468382/2019-00)]. 
The authors are grateful for the referees' comments 
which allow a significant improvement of the manuscript.
\end{acknowledgments}

\hfill


\appendix

\section{Proof of Lemma~\ref{Lemmadetc}} \label{AppLema1}
Let us prove first that $\DD>0$ under the hypothesis of the Lemma.
%
Since $\CC$ is antisymmetric ($\CC = - \CC^\tp$) and invertible, 
the spectral theorem \cite{Horn1990} guarantees that there is an unitary diagonalizing matrix $\UU$, such that 
$\UU \CC \UU^\dagger = (-\imath \mathbb{c}) \oplus (\imath\mathbb{c})=\CC^\prime$, 
where $\mathbb{c} = \mbox{diag}(c_1, \ldots, c_n)$, 
with $c_i>0$, for $i=1,\ldots,n$. 
Thus, using \eqref{DDplusiCC} we have $\UU\hbar \GM \UU^\dagger = 
\DD^\prime + \CC^\prime \ge 0$, where $\DD^\prime = \UU \DD \UU^\dagger$ and 
$\CC^\prime=(\hbar\mathbb{c}) \oplus (-\hbar\mathbb{c})$.
Now consider the matrix $\GM^\tp$ and apply the same unitary transformation, 
reaching $\UU \GM^\tp \UU^\dagger = \DD' - \CC' \ge 0$, 
which is positive-semidefinite, since $\GM^\tp$ is Hermitian and has the same 
eigenvalues of $\GM$. 
For a complex $2n$-dimensional vector $\bf z$, 
the above matrix inequalities are equivalent to 
${\bf z}^\dagger \DD' {\bf z} \pm {\bf z}^\dagger \CC' {\bf z} \ge 0$,
for all vectors ${\bf z}$, which implies ${\bf z}^\dagger \DD' {\bf z} > 0$, thus $\DD > 0$.  
\par 
To prove that $\GM>0$, it is enough to use that ${\bf z}^\dagger \CC {\bf z} = 0$ for all vectors $\bf z$, 
since $\CC$ is antisymmetric. 
Consequently, $\hbar {\bf z}^\dagger \GM {\bf z} = {\bf z}^\dagger \DD {\bf z}$, 
since $\DD > 0$ under the hypothesis of the in Sec.\ref{conclusions}, thus $\GM >0$. 
This finishes the proof.
 
Observing the particular case where $\CC = 0$, one has $\GM >0$ iff $\DD >0$, 
which shows that the converse of the Lemma is not true.

\section{Proof of Theorem~\ref{DBimpliesStat}}
\label{AppDBimpliesStat}

Since 
$\bar{\Lambda}_t$ is a \hyp{GDS} and satisfies a $\hrhobgth$-DBC, we have
\bea
\expval{\bar{\Lambda}_t[\hat A],\hat B}_\text{GNS}=\expval{\hat A,\bar{\Lambda}_t[\hat B]}_\text{GNS}.
\eea
Evaluating for $\hat A=\hat 1$, we have $\bar{\Lambda}_t[\hat 1]=\hat 1$ and thus
\beq
\expval{\hat 1,\hat B}_\text{GNS}=\expval{\hat 1,\bar{\Lambda}_t[\hat B]}_\text{GNS}.
\eeq
However,  
$\expval{\hat 1,\hat B}_\text{GNS}=\expval{\hrhobgt,\hat B}$ and
$\expval{\hat 1,\bar{\Lambda}_t[\hat B]}_\text{GNS}=\expval{\hrhobgt,\bar{\Lambda}_t[\hat B]}=
\expval{\Lambda_t[\hrhobgt],\hat B}$, where we use that $\Lambda_t$ is the adjoint 
of $\bar{\Lambda}_t$ with respect to the Hilbert-Schmidt scalar product $\expv{\cdot}$. 
Therefore, we arrive to 
\bea
\expval{\hrhobgth,\hat B}=
\expval{\Lambda_t[\hrhobgth],\hat B},
\label{Lambdativariante}
\eea
valid for all operators $\hat B$ such the Hilbert-Schmidt scalar product on both sides is finite.
In particular, we can use $\hat B=\hat R_{\x}$, a reflection operator defined in Section~\ref{WFFPE}, 
therefore, \eqref{Lambdativariante} it is just the equality between the Wigner symbols of 
$\hrhobgth$ and $\Lambda_t[\hrhobgth]$, which implies that $\Lambda_t[\hrhobgth]=\hrhobgth$.

\section{Proof of Theorem~\ref{GDScommod}}
\label{AppGDScommod}
Let us start writing
\bea
&&\expval{\bar{\Lambda}_t[\Xi_{\tau}[\hat A]],\hat B}_\text{GNS}=
\Tr(\hrhobgth (\bar{\Lambda}_t[\Xi_{\tau}[\hat A]])^\dagger\hat B ) \nonumber \\ 
&&=\Tr(\hrhobgth (\Xi_{\tau}[\hat A])^\dagger \bar{\Lambda}_t[\hat B] ) =
\Tr(\hrhobgth \Xi_{{\tau^\ast}}[\hat A^\dagger]\bar{\Lambda}_t[\hat B] ) \nonumber\\
&&=\Tr(\hrhobgth\,e^{\frac{\imath}{\hbar} \hH {\tau^\ast}}\,\hat A^\dagger e^{-\frac{\imath}{\hbar} \hH {\tau^\ast}}\, \bar{\Lambda}_t[\hat B])\nonumber \\
&&=\Tr(e^{-\frac{\imath}{\hbar} \hH {\tau^\ast}}\, \bar{\Lambda}_t[\hat B^\dagger]^\dagger\, e^{\frac{\imath}{\hbar} \hH {\tau^\ast}} \hrhobgth \hat A^\dagger ),
\label{aux1QNS}
\eea
where in the first line we used that $\bar{\Lambda}_t$ satisfies a $\hrhobgth$-{\hyp{DBC}} and the notation $\tau^*$ means complex conjugation of $\tau$;
in the third line, we used that $[e^{\pm\frac{\imath}{\hbar} \hH \tau},\hrhobgth]=0$ and 
that $\bar{\Lambda}_t$ is a real superoperator, {\it i.e.}, 
$\bar{\Lambda}_t[\hat B^\dagger]^\dagger=\bar{\Lambda}_t[\hat B]$.

Following \cite{Carlen2017}, 
we specialize~\eqref{aux1QNS} for ${\tau^\ast}=-\imath\hbar\beta$, 
therefore $e^{\frac{\imath}{\hbar} \hH (-\imath\hbar\beta)}=
(\hrhobgth)^{-1}$, then
\bea
&&\expval{\bar{\Lambda}_t[\Xi_{\imath\hbar\beta}[\hat A]],\hat B}_\text{GNS} = \Tr(\hrhobgth\, \bar{\Lambda}_t[\hat B^\dagger]^\dagger\, 
 \hat A^\dagger ) \nonumber \\
&&= \Tr(\hrhobgth\, (\hat B^\dagger)^\dagger\,\bar{\Lambda}_t[
 \hat A^\dagger] )=\Tr(\hrhobgth\, \hat B\,\bar{\Lambda}_t[
 \hat A]^\dagger) \nonumber \\
&& = \Tr(\bar{\Lambda}_t[
 \hat A]^\dagger\,\hrhobgth\, \hat B)
=\Tr(\hrhobgth\,
(\hrhobgth)^{-1}\bar{\Lambda}_t[ \hat A]^\dagger\,\hrhobgth\, \hat B) \nonumber\\
&&=\Tr(\hrhobgth\,
{(\hrhobgth\bar{\Lambda}_t[ \hat A]\, (\hrhobgth)^{-1})}^\dagger\, \hat B)\nonumber\\
&&= \Tr(\hrhobgth\,(\Xi_{\imath\hbar\beta}[\bar{\Lambda}_t[
 \hat A]])^\dagger\, \hat B) \nonumber\\
&&=
 \expval{\Xi_{\imath\hbar\beta}[\bar{\Lambda}_t[\hat A]],\hat B}_\text{GNS}.
 \label{commuteXiLamb}
\eea
Since this identity is valid for any operator $\hat B$, 
we can set it equal to the reflection operator, \ie
$\hat B=\hat R_{\x}$.  
Then, in this case, Eq.\eqref{commuteXiLamb} 
just restates that the Wigner symbols of the operator
$\hrhobgth\bar{\Lambda}_t[\Xi_{\imath\hbar\beta}[\hat A]]$ 
and of $\hrhobgth\Xi_{\imath\hbar\beta}[\bar{\Lambda}_t[\hat A]]$ are identical, 
therefore also the operators themselves.
For any finite value $\beta$, the operator $\hrhobgth$ has an inverse, then it must be true that 
\beq
\label{LambconmXi}
\bar{\Lambda}_t[\Xi_{\imath\hbar\beta}[\hat A]]=\Xi_{\imath\hbar\beta}[\bar{\Lambda}_t[\hat A]],
\eeq
for all $\hat A$, which means that $\bar{\Lambda}_t$ commutes with $\Xi_{\imath\hbar\beta}$ for all $t$.
Noting that 
$(\hrhobgth)^{\pm\frac{\imath t}{\hbar\beta}}=(e^{-\beta\hH})^{\pm\frac{\imath t}{\hbar\beta}}=
e^{{\mp}\frac{\imath}{\hbar}t\hH}$, therefore we can define 
$(\Xi_{\imath\hbar\beta})^{{-\frac{\imath t}{\hbar\beta}}}[\cdot]={(\hrhobgth)^{-\frac{\imath t}{\hbar\beta}}\,\cdot\,((\hrhobgth)^{-1})^{-\frac{\imath t}{\hbar\beta}}}= e^{\frac{\imath }{\hbar} \hH t}\,\cdot\,e^{-\frac{\imath}{\hbar} \hH t}=\Xi_t[\cdot]$.
But, according to \eqref{LambconmXi}, $\bar{\Lambda}_t$ commutes with $\Xi_{\imath\hbar\beta}$ 
and thus it must commute
with any function of $\Xi_{\imath\hbar\beta}$, then  
with $(\Xi_{\imath\hbar\beta})^{\frac{\imath}{\hbar}t}=\Xi_t$ for all $t\in\Cset$.
Taking the time derivative on both sides of Eq.\eqref{LambconmXi} and using  
$d\bar{\Lambda}_t[\cdot]/dt=\bar{\opL}[\bar{\Lambda}_t[\cdot]]$, 
we get 
$\bar{\opL}[\bar{\Lambda}_t[\Xi_{\imath\hbar\beta}[\hat A]]]=\Xi_{\imath\hbar\beta}[\bar{\opL}\bar{\Lambda}_t[[\hat A]]]$,
which shows that $\bar{\opL}$ commute with $\Xi_{\imath\hbar\beta}$ when $t=0$. 
Following the same reasoning as before,  
$\bar{\opL}$ commutes with $\Xi_t$.
%
\section{Demonstration of Eq.\eqref{invaderiva}} 
\label{Appinvaderiva}
Let us start from the left-hand side of~\eqref{invaderiva}:
\begin{widetext}
\bea
\Xi_{-t}\left[\pdv{}{\hbx^\tp}[\cdot]\right]&=&
\Xi_{-t}\left[\frac{\imath}{\hbar}[(\JJJ\hbx)^\tp,\cdot]\right]=\frac{\imath}{\hbar}\left(\Xi_{-t}\left[(\JJJ\hbx)^\tp\,\cdot\right] -\frac{\imath}{\hbar}\Xi_{-t}\left[\cdot\,(\JJJ\hbx)^\tp\right] \right)
=\frac{\imath}{\hbar}\left(\Xi_{-t}\left[(\JJJ\hbx)^\tp\right]\Xi_{-t}\left[\cdot\right] -
\frac{\imath}{\hbar}\Xi_{-t}\left[\cdot\right] \Xi_{-t}\left[(\JJJ\hbx)^\tp\right]\right)\nonumber\\
&=&
\frac{\imath}{\hbar}\left(-\Xi_{-t}\left[\hbx^\tp\right]\,\JJJ\,\Xi_{-t}\left[\cdot\right] +
\frac{\imath}{\hbar}\Xi_{-t}\left[\cdot\right]\,\Xi_{-t}\left[\hbx^\tp\right]\,\JJJ\right)=
\frac{\imath}{\hbar}\left(-\hbx^\tp \SST_t^{-\tp}\,\JJJ\,\Xi_{-t}\left[\cdot\right] +
\frac{\imath}{\hbar}\Xi_{-t}\left[\cdot\right]\,\hbx^\tp \SST_t^{-\tp}\,\JJJ\right)\nonumber\\
&=&
\frac{\imath}{\hbar}\left(-\hbx^\tp \,\JJJ\,\SST_t\,\Xi_{-t}\left[\cdot\right] +
\frac{\imath}{\hbar}\Xi_{-t}\left[\cdot\right]\,\hbx^\tp\,\JJJ\SST_t\right)=
\frac{\imath}{\hbar}\left((\JJJ\hbx)^\tp \Xi_{-t}\left[\cdot\right] -
\frac{\imath}{\hbar}\Xi_{-t}\left[\cdot\right]\,(\JJJ\,\hbx)^\tp\right)\SST_t=\pdv{}{\hbx^\tp}
\left[\Xi_{-t}[\cdot]\right]\SST_t,
\eea
\end{widetext}
where we used that $\Xi_{-t}[\hbx^\tp]=\hbx\,\SST_t^{-\tp}$ and  $\SST_t^{-\tp}\JJJ=\JJJ\SST_t$ 
because $\SST_t$ is a symplectic matrix.

\section{Demonstration of Eq.\eqref{coolformGM}} 
\label{AppcoolformGM}

In order to prove Eq.\eqref{coolformGM}, 
note that the decoherence matrix $\GM$, that stems from Theorem~\ref{TeocondDandC}, must be invariant under a 
congruence with $\SST_t$ in~\eqref{SSStdef}. 
This follows using \eqref{conginvDDandCC} into the definition of the matrices $\DD$ and $\CC$ in~\eqref{defMats}, and the fact that 
$\SST_t$ is a real matrix.
\par
Now, using Eq.\eqref{SSStdef} we can write 
$\SST_t(\QQ\SSSth)^{-1}=(\QQ\SSSth)^{-1}e^{(\JJJ\BB)_{\rm d}\,t}$, 
so the column vectors $(\QQ\SSSth)^{-1}|_k=\lb_k$ for $k=1,\ldots,2n$ 
are the complex eigenvectors of $\SST_t$. Respecting the block order in the matrix in~\eqref{SpecJB}, we can write
\bse
\label{eigenvecStilde}
\bea
\SST_t\,\lb_j&=&e^{\imath \omega_j t}\lb_j,\\
\SST_t\lb_{n+j}&=&\SST_t\lb_j^*=e^{-\imath \omega_j t}\lb_j^*,
\eea
\ese
where $j=1,\ldots,n$ and $\omega_j>0$ are the frequencies 
corresponding to the symplectic spectra of $\BB$ in~\eqref{diagsymplecBB}.
According to \eqref{eigenvecStilde}, 
it is clear that the canonical form of $\GM$ in \eqref{coolformGM} 
is manifestly invariant under a congruence through $\SST_t$ 
and is positive-definite, as it must be.  
\par
Let us now prove that the matrices $\ss$ and $\rr$ in~\eqref{coolformGM} 
satisfy the relation in Eq.\eqref{matbb}.
We first note that the expression of $\GM$ in \eqref{coolformGM} 
can be rewritten as in \eqref{rewriteGM}, which allows us to show that the matrix 
$(\SSSth)^{\tp}\QQ^\dagger$ diagonalizes simultaneously the Hamiltonian matrix 
$\JJJ\DD$ and the skew Hamiltonian matrix $\JJJ\CC$, 
therefore, we must have 
$[\JJJ\DD,\JJJ\CC]=0$.
Indeed, from \eqref{rewriteGM}, we have 
\bea
\JJJ\DD&=&\hbar\JJJ(\SSSth)^{-1}\Lambt_r(\SSSth)^{-\tp}=\hbar(\SSSth)^\tp\JJJ \Lambt_r(\SSSth)^{-\tp}\nonumber\\
&=&\hbar(\SSSth)^\tp\QQ^\dagger\QQ\JJJ \Lambt_r\QQ^\dagger\QQ(\SSSth)^{-\tp}\nonumber\\
&=&(\SSSth)^\tp\QQ^\dagger(\JJJ\DD)_{\rm d}((\SSSth)^\tp\QQ^\dagger)^{-1},
\label{JDdiagonalized}
\eea
where $\QQ$ is the unitary matrix defined in~\eqref{defQQ}. 
In the above steps, the symplectic condition $\JJJ(\SSSth)^{-1}=(\SSSth)^\tp\JJJ$ 
and the fact that $\hbar\QQ\JJJ \Lambt_r\QQ^\dagger$ is a diagonal matrix were employed.
%
Since Eq.(\ref{JDdiagonalized}) is a similarity transformation 
the matrix $\hbar\QQ\JJJ \Lambt_r\QQ^\dagger$ has 
the eigenvalues of $\JJJ\DD$ in its diagonal.
Therefore, we can write
\beq
(\JJJ\DD)_{\rm d}=\frac{\imath\hbar}{2}  \left( (|\ss|^2+|\rr|^2) \oplus  \left(-|\ss|^2-|\rr|^2\right)\right).
\eeq
It is worth to note that, according to Lemma~\ref{LemmaJOdiag},
the diagonal matrix $\dd =\tfrac{1}{2}(|\ss|^2+|\rr|^2)$ 
contains the symplectic spectrum of $\DD$.
In an analogous way, we have
\bea
\JJJ\CC&=&\JJJ(\SSSth)^{-1}(-\JJJ\Lambt_i)(\SSSth)^{-\tp}=(\SSSth)^\tp \Lambt_i(\SSSth)^{-\tp} \nonumber\\
&=&(\SSSth)^\tp\QQ^\dagger\Lambt_i\QQ(\SSSth)^{-\tp}\nonumber\\
&=&(\SSSth)^\tp\QQ^\dagger(\JJJ\CC)_{\rm d}((\SSSth)^\tp\QQ^\dagger)^{-1},
\label{JCdiagonalized}
\eea
where the eigenvalue matrix corresponding to $\JJJ\CC$ is 
\beq
(\JJJ\CC)_{\rm d}=\QQ^\dagger\Lambt_i\QQ=\Lambt_i  = \jc.
\eeq
From~\eqref{JDdiagonalized} and~\eqref{JCdiagonalized}, 
we immediately realize that $\JJJ\DD$ commutes with 
$\JJJ\CC$, as we wanted to prove.
\par
The condition $[\JJJ\DD,\JJJ\CC]=0$ in Theorem~\ref{theoremvNUzero} (proved in the main text) 
determines univocally the covariance matrix $\VVth$
of a thermal equilibrium state $\hrhog^{\mathrm{th}}$. 
Therefore, using the expressions ~\eqref{Vthposta} and ~\eqref{DDandCC-DBC}, 
we can write 
\beq
\VVth=(\SSSth)^{-1}\frac{1}{2}\Lambt_r\Lambt_i^{-1}(\SSSth)^{-\tp},
\label{diagsympVVth}
\eeq
where we used the symplectic condition $\JJJ(\SSSth)=(\SSSth)^{-\tp}\JJJ$.
Comparing with Eq.\eqref{VVthsymdiag}, 
we immediately recognize  in Eq.\eqref{diagsympVVth} 
the symplectic spectrum of $\VVth$, \ie 
\beq
\frac{1}{2}\Lambt_r\Lambt_i^{-1}=\kkth\oplus\kkth.
\label{kthexpression}
\eeq 
Equating the matrix elements on both sides of~\eqref{kthexpression}, 
we obtain $\frac{1}{2}\frac{1+a_j}{1-a_j}=\kappa_j^{\rm th}$ with $a_j=\frac{|r_j|^2}{|s_j|^2}$ and $j=1,\ldots,n$. Inverting these equations and rewriting in a matrix structure, we get
\beq
\frac{|\rr|^2}{|\ss|^2}=\frac{2\kkth-\In}{2\kkth+\In}.
\label{rjsjrelkkth}
\eeq
Now using~\eqref{symplectdiagUU}, ~\eqref{UUth}, and~\eqref{diagsymplecBB}, 
we arrive to the equality $g(\kkth)=\hbar \beta \ww$. Then, employing the identity 
$\exp[{-2\coth^{-1}(2x)}] = \frac{2x-1}{2x+1}$ for $x\ge \frac{1}{2}$, 
we can rewrite Eq.\eqref{rjsjrelkkth} as
\beq
\label{rr2sobress2}
\frac{|\rr|^2}{|\ss|^2}=e^{-\hbar\beta \ww},
\eeq 
which finally implies the relation in~\eqref{matbb}.

\section{Proof of Eq.\eqref{commhXmema}}
\label{AppLindQDBC}

Let us prove an equivalent statement of Eq.\eqref{commhXmema}, \ie in a \hyp{GDS} satisfying a $\hrhobgth$-{\rm DBC}
we have
\beq
\label{commHLj}
[\hH,\hL_j]=-\hbar\omega_j\hL_j\Rightarrow [\hH,\hL^\dagger_j]=\hbar\omega_j\hL^\dagger_j,
\eeq
for $j=1,\ldots,n$, $\hH$ given in~\eqref{Hform}, 
and the Lindblad operators $\hL_j$ in~\eqref{LindbopGDSwDBC}.

For a generic symmetric matrix $\mathbb X$ and a generic complex vector $\l$, 
it is straightforward to show that 
\beq
\label{LjGDSDBC}
[\tfrac{1}{2}\hbx^\tp \mathbb X  \hbx,\l^\tp\JJJ\hbx]=\imath\hbar\l^\tp \mathbb X \hbx=
\imath\hbar\l^\tp\mathbb X \JJJ^\tp\JJJ\hbx,
\eeq 
which is the commutation relation for a generic quadratic Hamiltonian 
with a linear Lindblad operator. 
But, if the \hyp{GDS} satisfies a $\hrhobgth$-\hyp{DBC} according to Eq.\eqref{SSStdef}, $\lb_j$ appearing in \eqref{defhXmahXme} 
is an eigenvector of the matrix $\JJJ\BB$, \ie $\JJJ\BB\lb_j=\imath\omega_j \lb_j$, 
which is the same as say that $\lb_j=(\QQ\SSSth)^{-1}|_j$, see Appendix~\ref{AppcoolformGM}. 
Therefore, using (\ref{LjGDSDBC}) with $\l_j^\tp\BB\JJJ=\imath\omega_j\l_j^\tp$ 
we arrive to~\eqref{commHLj}.\\

\section{Proof of Eq.\eqref{si2value}}
\label{Appmatss}

Let us start noting that for the thermal state $\hrhog^{\mathrm{th}}$ in~\eqref{GibbsstateH}, 
or equivalently in~\eqref{hrhogdeco}, the eigenenergy 
distribution is the usual (multimode) Planck distribution:
\bea
\!\! P({\bf n})=\ev{\hrhog^{\mathrm{th}}}{\phi_{\bf n}}=\ev{\hrhog_{\rm ho}^{\mathrm{th}}}{{\bf n}} 
= \prod_{j=1}^n P_{n_{j}}(\bar{n}_j), 
\eea
which is nothing but a consequence of Eqs.(\ref{hHhHho}) and (\ref{eigenvhH}). 
In the above equation, 
\bea\label{Boltzdisti1}
P_{n_{j}}(\bar{n}_j) = \frac{1}{\bar{n}_j + 1} \left(\frac{\bar{n}_j}{\bar{n}_j + 1}\right)^{n_j}, 
\eea
where $\bar{n}_j$ is in~\eqref{Planckfactor}. 
Note that $ \bar{n}_j(\bar{n}_j+1)^{-1}=e^{-\hbar\beta\omega_j}$.
\par
This distribution can be recovered considering the stationary 
regime $\dv{\hrho_t}{t} = 0$ in~\eqref{MEGDSwDBC}, 
subsequently taking the diagonal matrix elements in the eigenbasis \{$\ket{\phi_{\bf n}}\}$ 
and using the commutation relation $[\hXme_j,\hXma_j]=1$, indeed 
\bea
&&\sum_{j=1}^n \frac{|s_j|^2}{\hbar}  \left((n_j+1) \,{P}_{n{_j}+1}-n_j{P}_{n{_j}} \right) + \nonumber \\
&&\sum_{j=1}^n \frac{|s_j|^2}{\hbar} e^{-\hbar\beta\omega_j} \left(n_j\;{P}_{n{_j}-1}-(n_j+1){P}_{n{_j}}\right) =0, \nonumber
\eea
where ${P}_{n{_j}} = \langle \phi_{\bf n}|  \hrhog^{\mathrm{th}} |\phi_{\bf n}\rangle$ 
and ${P}_{n{_j}\pm 1} = \langle \phi_{{\bf n}_j^{\pm}} |\hrhog^{\mathrm{th}}|\phi_{{\bf n}_j^\pm}\rangle$. As can be directly checked, the solution is  
\beq
\label{Boltzdisti2}
{P}_{n{_j}} = \frac{\hbar\bar{\gamma}_j}{|s_j|^2}e^{-\hbar\beta\omega_jn_j}.
\eeq
The constant $\bar{\gamma}_j$ is included in order to ${\hbar\bar{\gamma}_j}/{|s_j|^2}$ be dimensionless. 
Comparing Eqs.~\eqref{Boltzdisti2} and~\eqref{Boltzdisti1} we arrive to~\eqref{si2value}. 
\par
\section{The algebra of commuting elliptic Hamiltonian matrices}
\label{AppComRel}
In this appendix we will prove the following theorem:
\begin{theorem}
Consider a positive definite symmetric matrix $\BB$ and a symmetric matrix $\BBp$ 
such that $[\JJJ\BB,\JJJ\BBp]=0$, then 
\beq
\label{algebraJBp}
\JJJ\BBp=(\QQ\SSS)^{-1} \,\imath \xx \oplus (-\imath\xx)\, (\QQ\SSS),
\eeq
where $\QQ$ is defined in (\ref{defQQ}), $\xx:={\rm{diag}}(\lambda_1,\ldots,\lambda_n)$ with 
$\lambda_i \in \Rset$ for $i=1,\ldots,n$, 
and $\SSS$ is a symplectic matrix that diagonalizes symplectically $\BB$, \ie
\beq\label{algebraJBp2}
\SSS^{-\tp}\BB \SSS= \ww \oplus \ww, \,\,\, \ww={\rm{diag}}(\omega_1,\ldots,\omega_n) 
\eeq
for $\omega_i>0$ and $i=1,\ldots,n$. 
Further, Eq.(\ref{algebraJBp}) is equivalent to a ``symplectic diagonalization'' 
of $\BBp$: 
\beq \label{algebraJBp2a}
\SSS^{-\tp}\BBp\SSS^{-1}=\xx\oplus\xx.
\eeq
\end{theorem}
\par
Let $\SSSp$ be a symplectic matrix such that $\SSSp^{-\tp}\BB \SSSp= \ww \oplus \ww$,
then 
\bea
[\JJJ\BB,\JJJ\BBp] &=& [\JJJ {\SSSp}^{\tp}(\ww\oplus\ww) \SSSp, \JJJ \BBp] \nonumber\\
&=& \JJJ \SSSp^{\tp} \, [ \JJJ(\ww\oplus\ww) , {\SSSp}^{-\tp} \BBp {\SSSp}^{-1}] \, {\SSSp}, 
\label{commtrel2}
\eea
where we used the symplectic condition 
${\SSSp} \JJJ = \JJJ {\SSSp}^{-\tp}$ and  
$\JJJ(\ww\oplus\ww) = (\ww\oplus\ww)\JJJ$.  
Writing 
\[
\tilde{\BB} := 
{\SSSp}^{-\tp} \BBp {\SSSp}^{-1} = 
\begin{pmatrix} 
       \tilde{\xx} &  \tilde{\yy}  \\
       \tilde{\yy}^{\tp} &   \tilde{\zz} \\
   \end{pmatrix}, 
\]
where  $\tilde{\xx}, \tilde{\yy}$ and $\tilde{\zz}$ are  $n\times n$ 
real matrices such that $\tilde{\xx} = \tilde{\xx}^\tp$ and $\mathbb z = \mathbb z^\tp$, 
and using the commutation relation $[\JJJ\BB,\JJJ\BBp]=0$ in~\eqref{commtrel2} one attains
\[
\tilde{\yy}^\top  = -\ww^{-1} \tilde{\yy} \ww = -\ww \tilde{\yy} \ww^{-1} \,\,\, {\text{and}} \,\,\, 
\tilde{\zz}  = \ww^{-1} \tilde{\xx} \ww = \ww \tilde{\xx} \ww^{-1};
\]
consequently, $[\tilde{\yy}, \ww] = [\tilde{\xx}, \ww] = 0$, 
thus $\tilde{\xx} = \tilde{\zz}$ and 
$\tilde{\yy}$ is skew-symmetric $\tilde{\yy}^\tp = - \tilde{\yy}$.
All of these relations enable us to write 
\beq
\tilde{\BB} := 
{\SSSp}^{-\tp} \BBp {\SSSp}^{-1} = 
\begin{pmatrix} 
      \tilde{\xx} & \tilde{\yy}  \\
      -\tilde{\yy} & \tilde{\xx} \\
   \end{pmatrix}.    
   \label{Btilde}
\eeq
Multiplying by $\JJJ$ from left, considering the symplecticity of $\SSSp$, and using $\QQ$, 
last equation is equivalently rewritten as  
\beq
(\QQ\SSSp)\, \JJJ\BBp \,(\QQ\SSSp)^{-1} = \QQ \JJJ \tilde{\BB} \QQ^\dag = 
(\imath \tilde{\xx} - \tilde{\yy}) \oplus (-\imath \tilde{\xx} - \tilde{\yy}).     
\label{QQSSpJBp}
\eeq
The above particular block structure is a consequence of 
the degenerated structure of the diagonal matrix $\ww\oplus\ww$, 
where each diagonal element is at least doubly-degenerated.
\par
The two blocks in the matrix of rightmost equality in~\eqref{QQSSpJBp} are skew-Hermitian 
and moreover they are complex conjugate of each other.
Recalling that a skew-Hermitian matrix has pure imaginary (possibly null) 
eigenvalues and is unitarily diagonalizable \cite{Horn2013}, 
then there exists an unitary matrix $\uu$ such that 
$\uu(\imath \tilde{\xx} - \tilde{\yy})\uu^\dag = \imath \xx$, 
where $\xx=\text{diag}(\lambda_1,\ldots,\lambda_n)$
is the diagonal matrix containing the eigenvalues $\lambda_j \in \Rset$ 
of the matrix $\imath \tilde{\xx}-\tilde{\yy}$.
\par
Defining $\SSS=\RRR\,\SSSp$ with 
\beq
\RRR=\QQ
   \begin{pmatrix} 
      \uu & 0 \\
      0& \uu^* \\
   \end{pmatrix}\QQ^{\dagger} \,\, \in {\rm Sp}(2n,\Rset)\;\cap\;{\rm O}(2n)
\eeq
a real symplectic and orthogonal matrix, from (\ref{QQSSpJBp}) we can write
\beq
\label{algebraJBp3}
\QQ\SSS\;\JJJ\BBp\; (\QQ\SSS)^{-1}= \,\imath \xx \oplus (-\imath\xx),
\eeq
which is Eq.(\ref{algebraJBp}). Since $\SSS$ is symplectic and using that $\QQ^\dagger \,(\imath \xx \oplus (-\imath\xx))\QQ =\JJJ (\xx\oplus\xx) $, we immediately recover Eq.\eqref{algebraJBp2a} from~\eqref{algebraJBp3}.
\par
However it remains to prove that $\SSS$ satisfies (\ref{algebraJBp2}).
If the symplectic spectrum in $\ww$ is non-degenerate,  
conditions $[\tilde{\yy}, \ww] = [\tilde{\xx}, \ww] = 0$, $\tilde{\xx}^\tp=\tilde{\xx}$, and $\tilde{\yy}^\tp=-\tilde{\yy}$ 
imply $\tilde{\xx}=\xx=\text{diag}(\lambda_1,\ldots,\lambda_n)$ and $\yy=0$, so one can choose $\uu=\In$ 
such that $\RRR=\In$ and $\SSS=\SSSp$; 
consequently $\SSS$ satisfies (\ref{algebraJBp2}), as claimed.
When the symplectic spectrum in $\ww$ is degenerate, 
$\tilde{\xx}$ is diagonal  outside the degenerate subspaces of $\ww$, while
$\yy$ is null outside the same subspaces. Therefore, 
the unitary matrix $\uu$ have to diagonalize $\imath \tilde{\xx}-\tilde{\yy}$ only inside the degenerate subspaces of $\ww$.
This is possible choosing $\uu$ block diagonal such  $\uu\imath \tilde{\xx}-\tilde{\yy}\uu^\dagger=\imath \xx$. In this case 
we also have that $\RRR^\tp\ww\oplus\ww\,\RRR=\ww\oplus\ww$, and
therefore $\SSS=\RRR\,\SSSp$ also diagonalize symplectically $\BB$,
\ie
\beq
\BB=\SSS^{\tp}\ww\oplus\ww\SSS=\SSSp^{\tp}\RRR^\tp\ww\oplus\ww\,\RRR\SSSp=
\SSSp^{\tp}\ww\oplus\ww\,\SSSp, 
\eeq
which shows that $\SSS$ satisfies (\ref{algebraJBp2}) for the degenerate case, with this we finish the proof of the theorem.
%


\begin{thebibliography}{46}%
\makeatletter
\providecommand \@ifxundefined [1]{%
 \@ifx{#1\undefined}
}%
\providecommand \@ifnum [1]{%
 \ifnum #1\expandafter \@firstoftwo
 \else \expandafter \@secondoftwo
 \fi
}%
\providecommand \@ifx [1]{%
 \ifx #1\expandafter \@firstoftwo
 \else \expandafter \@secondoftwo
 \fi
}%
\providecommand \natexlab [1]{#1}%
\providecommand \enquote  [1]{``#1''}%
\providecommand \bibnamefont  [1]{#1}%
\providecommand \bibfnamefont [1]{#1}%
\providecommand \citenamefont [1]{#1}%
\providecommand \href@noop [0]{\@secondoftwo}%
\providecommand \href [0]{\begingroup \@sanitize@url \@href}%
\providecommand \@href[1]{\@@startlink{#1}\@@href}%
\providecommand \@@href[1]{\endgroup#1\@@endlink}%
\providecommand \@sanitize@url [0]{\catcode `\\12\catcode `\$12\catcode
  `\&12\catcode `\#12\catcode `\^12\catcode `\_12\catcode `\%12\relax}%
\providecommand \@@startlink[1]{}%
\providecommand \@@endlink[0]{}%
\providecommand \url  [0]{\begingroup\@sanitize@url \@url }%
\providecommand \@url [1]{\endgroup\@href {#1}{\urlprefix }}%
\providecommand \urlprefix  [0]{URL }%
\providecommand \Eprint [0]{\href }%
\providecommand \doibase [0]{https://doi.org/}%
\providecommand \selectlanguage [0]{\@gobble}%
\providecommand \bibinfo  [0]{\@secondoftwo}%
\providecommand \bibfield  [0]{\@secondoftwo}%
\providecommand \translation [1]{[#1]}%
\providecommand \BibitemOpen [0]{}%
\providecommand \bibitemStop [0]{}%
\providecommand \bibitemNoStop [0]{.\EOS\space}%
\providecommand \EOS [0]{\spacefactor3000\relax}%
\providecommand \BibitemShut  [1]{\csname bibitem#1\endcsname}%
\let\auto@bib@innerbib\@empty
\bibitem [{\citenamefont {Holevo}\  {et~al.}(1999)\citenamefont {Holevo},
  \citenamefont {Sohma},\ and\ \citenamefont {Hirota}}]{Holevo1999}%
  \BibitemOpen
  \bibfield  {author} {\bibinfo {author} {\bibfnamefont {A.~S.}\ \bibnamefont
  {Holevo}}, \bibinfo {author} {\bibfnamefont {M.}~\bibnamefont {Sohma}},\ and\
  \bibinfo {author} {\bibfnamefont {O.}~\bibnamefont {Hirota}},\ }\href
  {https://doi.org/10.1103/PhysRevA.59.1820} {\bibfield  {journal} {\bibinfo
  {journal} {Phys. Rev. A}\ }\textbf {\bibinfo {volume} {59}},\ \bibinfo
  {pages} {1820} (\bibinfo {year} {1999})}\BibitemShut {NoStop}%
\bibitem [{\citenamefont {Holevo}\ and\ \citenamefont
  {Werner}(2001)}]{Holevo2001}%
  \BibitemOpen
  \bibfield  {author} {\bibinfo {author} {\bibfnamefont {A.~S.}\ \bibnamefont
  {Holevo}}\ and\ \bibinfo {author} {\bibfnamefont {R.~F.}\ \bibnamefont
  {Werner}},\ }\href {https://doi.org/10.1103/PhysRevA.63.032312} {\bibfield
  {journal} {\bibinfo  {journal} {Phys. Rev. A}\ }\textbf {\bibinfo {volume}
  {63}},\ \bibinfo {pages} {032312} (\bibinfo {year} {2001})}\BibitemShut
  {NoStop}%
\bibitem [{\citenamefont {Holevo}(2002)}]{Holevo2002}%
  \BibitemOpen
  \bibfield  {author} {\bibinfo {author} {\bibfnamefont {A.~S.}\ \bibnamefont
  {Holevo}},\ }\bibinfo {title} {Sending quantum information with gaussian
  states},\ in\ \href {https://doi.org/10.1007/0-306-47097-7_10} {
  {\bibinfo {booktitle} {Quantum Communication, Computing, and Measurement
  2}}},\ \bibinfo {editor} {edited by\ \bibinfo {editor} {\bibfnamefont
  {P.}~\bibnamefont {Kumar}}, \bibinfo {editor} {\bibfnamefont {G.~M.}\
  \bibnamefont {D’Ariano}},\ and\ \bibinfo {editor} {\bibfnamefont
  {O.}~\bibnamefont {Hirota}}}\ (\bibinfo  {publisher} {Springer US},\ \bibinfo
  {address} {Boston, MA},\ \bibinfo {year} {2002})\ p.\ \bibinfo {pages}
  {75–82}\BibitemShut {NoStop}%
\bibitem [{\citenamefont {Cerf}\  {et~al.}(2007)\citenamefont {Cerf},
  \citenamefont {Leuchs},\ and\ \citenamefont {Polzik}}]{Cerf-book2007}%
  \BibitemOpen
  \bibfield  {author} {\bibinfo {author} {\bibfnamefont {N.~J.}\ \bibnamefont
  {Cerf}}, \bibinfo {author} {\bibfnamefont {G.}~\bibnamefont {Leuchs}},\ and\
  \bibinfo {author} {\bibfnamefont {E.~S.}\ \bibnamefont {Polzik}},\ }\href
  {https://doi.org/10.1142/p489} { {\bibinfo {title} {Quantum Information
  with Continuous Variables of Atoms and Light}}}\ (\bibinfo  {publisher}
  {Imperial College Press, London},\ \bibinfo {year} {2007})\ 
  \BibitemShut {NoStop}%
\bibitem [{\citenamefont {Caruso}\  {et~al.}(2008)\citenamefont {Caruso},
  \citenamefont {Eisert}, \citenamefont {Giovannetti},\ and\ \citenamefont
  {Holevo}}]{Caruso2008}%
  \BibitemOpen
  \bibfield  {author} {\bibinfo {author} {\bibfnamefont {F.}~\bibnamefont
  {Caruso}}, \bibinfo {author} {\bibfnamefont {J.}~\bibnamefont {Eisert}},
  \bibinfo {author} {\bibfnamefont {V.}~\bibnamefont {Giovannetti}},\ and\
  \bibinfo {author} {\bibfnamefont {A.~S.}\ \bibnamefont {Holevo}},\ }\href
  {https://doi.org/10.1088/1367-2630/10/8/083030} {\bibfield  {journal}
  {\bibinfo  {journal} {New Journal of Physics}\ }\textbf {\bibinfo {volume}
  {10}},\ \bibinfo {pages} {083030} (\bibinfo {year} {2008})}\BibitemShut
  {NoStop}%
\bibitem [{\citenamefont {Holevo}(2019)}]{Holevo2019}%
  \BibitemOpen
  \bibfield  {author} {\bibinfo {author} {\bibfnamefont {A.~S.}\ \bibnamefont
  {Holevo}},\ }\href {https://doi.org/10.1515/9783110642490} { {\bibinfo
  {title} {Quantum Systems, Channels, Information}}}\ (\bibinfo  {publisher}
  {De Gruyter},\ \bibinfo {year} {2019})\BibitemShut {NoStop}%
\bibitem [{\citenamefont {Weedbrook}\  {et~al.}(2012)\citenamefont
  {Weedbrook}, \citenamefont {Pirandola}, \citenamefont
  {Garc{\'i}a-Patr{\'o}n}, \citenamefont {Cerf}, \citenamefont {Ralph},
  \citenamefont {Shapiro},\ and\ \citenamefont {Lloyd}}]{Weedbrook2012}%
  \BibitemOpen
  \bibfield  {author} {\bibinfo {author} {\bibfnamefont {C.}~\bibnamefont
  {Weedbrook}}, \bibinfo {author} {\bibfnamefont {S.}~\bibnamefont
  {Pirandola}}, \bibinfo {author} {\bibfnamefont {R.}~\bibnamefont
  {Garc{\'i}a-Patr{\'o}n}}, \bibinfo {author} {\bibfnamefont {N.~J.}\
  \bibnamefont {Cerf}}, \bibinfo {author} {\bibfnamefont {T.~C.}\ \bibnamefont
  {Ralph}}, \bibinfo {author} {\bibfnamefont {J.~H.}\ \bibnamefont {Shapiro}},\
  and\ \bibinfo {author} {\bibfnamefont {S.}~\bibnamefont {Lloyd}},\ }\href
  {https://doi.org/10.1103/RevModPhys.84.621} {\bibfield  {journal} {\bibinfo
  {journal} {Reviews of Modern Physics}\ }\textbf {\bibinfo {volume} {84}},\
  \bibinfo {pages} {621} (\bibinfo {year} {2012})}\BibitemShut {NoStop}%
\bibitem [{\citenamefont {Adesso}\  {et~al.}(2014)\citenamefont {Adesso},
  \citenamefont {Ragy},\ and\ \citenamefont {Lee}}]{Adesso2014}%
  \BibitemOpen
  \bibfield  {author} {\bibinfo {author} {\bibfnamefont {G.}~\bibnamefont
  {Adesso}}, \bibinfo {author} {\bibfnamefont {S.}~\bibnamefont {Ragy}},\ and\
  \bibinfo {author} {\bibfnamefont {A.~R.}\ \bibnamefont {Lee}},\ }\href
  {https://doi.org/10.1142/S1230161214030012} {\bibfield  {journal} {\bibinfo
  {journal} {Open Systems {\&} Information Dynamics}\ }\textbf {\bibinfo
  {volume} {21}},\ \bibinfo {pages} {1440001} (\bibinfo {year}
  {2014})}\BibitemShut {NoStop}%
\bibitem [{\citenamefont {Wolf}\  {et~al.}(2006)\citenamefont {Wolf},
  \citenamefont {Giedke},\ and\ \citenamefont {Cirac}}]{Wolf2006}%
  \BibitemOpen
  \bibfield  {author} {\bibinfo {author} {\bibfnamefont {M.~M.}\ \bibnamefont
  {Wolf}}, \bibinfo {author} {\bibfnamefont {G.}~\bibnamefont {Giedke}},\ and\
  \bibinfo {author} {\bibfnamefont {J.~I.}\ \bibnamefont {Cirac}},\ }\href
  {https://doi.org/10.1103/PhysRevLett.96.080502} {\bibfield  {journal}
  {\bibinfo  {journal} {Physical Review Letters}\ }\textbf {\bibinfo {volume}
  {96}},\ \bibinfo {pages} {080502} (\bibinfo {year} {2006})}\BibitemShut
  {NoStop}%
\bibitem [{\citenamefont {Heinosaari}\  {et~al.}(2010)\citenamefont
  {Heinosaari}, \citenamefont {Holevo},\ and\ \citenamefont
  {Wolf}}]{Heinosaari2010}%
  \BibitemOpen
  \bibfield  {author} {\bibinfo {author} {\bibfnamefont {T.}~\bibnamefont
  {Heinosaari}}, \bibinfo {author} {\bibfnamefont {A.}~\bibnamefont {Holevo}},\
  and\ \bibinfo {author} {\bibfnamefont {M.}~\bibnamefont {Wolf}},\ }\href
  {https://doi.org/10.26421/QIC10.7-8-4} {\bibfield  {journal} {\bibinfo
  {journal} {Quantum Information and Computation}\ }\textbf {\bibinfo {volume}
  {10}},\ \bibinfo {pages} {619–635} (\bibinfo {year} {2010})}\BibitemShut
  {NoStop}%
\bibitem [{\citenamefont {Toscano}\  {et~al.}(2021)\citenamefont
  {Toscano}, \citenamefont {Bosyk}, \citenamefont {Zozor},\ and\ \citenamefont
  {Portesi}}]{Toscano2021}%
  \BibitemOpen
  \bibfield  {author} {\bibinfo {author} {\bibfnamefont {F.}~\bibnamefont
  {Toscano}}, \bibinfo {author} {\bibfnamefont {G.~M.}\ \bibnamefont {Bosyk}},
  \bibinfo {author} {\bibfnamefont {S.}~\bibnamefont {Zozor}},\ and\ \bibinfo
  {author} {\bibfnamefont {M.}~\bibnamefont {Portesi}},\ }\href
  {https://doi.org/10.1103/PhysRevA.104.062207} {\bibfield  {journal} {\bibinfo
   {journal} {Phys. Rev. A}\ }\textbf {\bibinfo {volume} {104}},\ \bibinfo
  {pages} {062207} (\bibinfo {year} {2021})}\BibitemShut {NoStop}%
\bibitem [{\citenamefont {Giovannetti}\  {et~al.}(2010)\citenamefont
  {Giovannetti}, \citenamefont {Holevo}, \citenamefont {Lloyd},\ and\
  \citenamefont {Maccone}}]{Giovannetti2010}%
  \BibitemOpen
  \bibfield  {author} {\bibinfo {author} {\bibfnamefont {V.}~\bibnamefont
  {Giovannetti}}, \bibinfo {author} {\bibfnamefont {A.~S.}\ \bibnamefont
  {Holevo}}, \bibinfo {author} {\bibfnamefont {S.}~\bibnamefont {Lloyd}},\ and\
  \bibinfo {author} {\bibfnamefont {L.}~\bibnamefont {Maccone}},\ }\href
  {https://doi.org/10.1088/1751-8113/43/41/415305} {\bibfield  {journal}
  {\bibinfo  {journal} {Journal of Physics A}\ }\textbf {\bibinfo {volume}
  {43}},\ \bibinfo {pages} {415305} (\bibinfo {year} {2010})}\BibitemShut
  {NoStop}%
\bibitem [{\citenamefont {{De P}alma}\  {et~al.}(2016)\citenamefont {{De
  P}alma}, \citenamefont {Trevisan},\ and\ \citenamefont
  {Giovannetti}}]{DePalma2016}%
  \BibitemOpen
  \bibfield  {author} {\bibinfo {author} {\bibfnamefont {G.}~\bibnamefont {{De
  P}alma}}, \bibinfo {author} {\bibfnamefont {D.}~\bibnamefont {Trevisan}},\
  and\ \bibinfo {author} {\bibfnamefont {V.}~\bibnamefont {Giovannetti}},\
  }\href {https://doi.org/10.1109/TIT.2016.2547426} {\bibfield  {journal}
  {\bibinfo  {journal} {IEEE Transactions on Information Theory}\ }\textbf
  {\bibinfo {volume} {62}},\ \bibinfo {pages} {2895} (\bibinfo {year}
  {2016})}\BibitemShut {NoStop}%
\bibitem [{\citenamefont {Alicki}\ and\ \citenamefont
  {Lendi}(2007)}]{Alicki2007}%
  \BibitemOpen
  \bibfield  {author} {\bibinfo {author} {\bibfnamefont {R.}~\bibnamefont
  {Alicki}}\ and\ \bibinfo {author} {\bibfnamefont {K.}~\bibnamefont {Lendi}},\
  }\href@noop {} { {\bibinfo {title} {{Quantum dynamical semigroups and
  applications}}}},\ Lecture notes in physics\ (\bibinfo  {publisher}
  {Springer},\ \bibinfo {address} {Berlin},\ \bibinfo {year}
  {2007})\BibitemShut {NoStop}%
\bibitem [{\citenamefont {Vanheuverzwijn}(1978)}]{Vanheuverzwijn1978}%
  \BibitemOpen
  \bibfield  {author} {\bibinfo {author} {\bibfnamefont {P.}~\bibnamefont
  {Vanheuverzwijn}},\ }\href
  {http://www.numdam.org/item/AIHPA_1978__29_1_123_0/} {\bibfield  {journal}
  {\bibinfo  {journal} {Annales de l'I.H.P. Physique th\'eorique}\ }\textbf
  {\bibinfo {volume} {29}},\ \bibinfo {pages} {123} (\bibinfo {year}
  {1978})}\BibitemShut {NoStop}%
\bibitem [{\citenamefont {Vanheuverzwijn}(1979)}]{Vanheuverzwijn1978errata}%
  \BibitemOpen
  \bibfield  {author} {\bibinfo {author} {\bibfnamefont {P.}~\bibnamefont
  {Vanheuverzwijn}},\ }\href
  {http://www.numdam.org/item/AIHPA_1979__30_1_83_0/} {\bibfield  {journal}
  {\bibinfo  {journal} {Annales de l'I.H.P. Physique th\'eorique}\ }\textbf
  {\bibinfo {volume} {30}},\ \bibinfo {pages} {83} (\bibinfo {year}
  {1979})}\BibitemShut {NoStop}%
\bibitem [{\citenamefont {Demoen}\  {et~al.}(1979)\citenamefont {Demoen},
  \citenamefont {Vanheuverzwijn},\ and\ \citenamefont {Verbeure}}]{Demoen1979}%
  \BibitemOpen
  \bibfield  {author} {\bibinfo {author} {\bibfnamefont {B.}~\bibnamefont
  {Demoen}}, \bibinfo {author} {\bibfnamefont {P.}~\bibnamefont
  {Vanheuverzwijn}},\ and\ \bibinfo {author} {\bibfnamefont {A.}~\bibnamefont
  {Verbeure}},\ }\href {https://doi.org/10.1016/0034-4877(79)90049-1}
  {\bibfield  {journal} {\bibinfo  {journal} {Reports on Mathematical Physics}\
  }\textbf {\bibinfo {volume} {15}},\ \bibinfo {pages} {27–39} (\bibinfo
  {year} {1979})}\BibitemShut {NoStop}%
\bibitem [{\citenamefont {S{\v{a}}ndulescu}\ and\ \citenamefont
  {Scutaru}(1987)}]{Sandulescu1987}%
  \BibitemOpen
  \bibfield  {author} {\bibinfo {author} {\bibfnamefont {A.}~\bibnamefont
  {S{\v{a}}ndulescu}}\ and\ \bibinfo {author} {\bibfnamefont {H.}~\bibnamefont
  {Scutaru}},\ }\href {https://doi.org/10.1016/0003-4916(87)90162-X} {\bibfield
   {journal} {\bibinfo  {journal} {Annals of Physics}\ }\textbf {\bibinfo
  {volume} {173}},\ \bibinfo {pages} {277} (\bibinfo {year}
  {1987})}\BibitemShut {NoStop}%
\bibitem [{\citenamefont {Carlen}\ and\ \citenamefont
  {Maas}(2017)}]{Carlen2017}%
  \BibitemOpen
  \bibfield  {author} {\bibinfo {author} {\bibfnamefont {E.~A.}\ \bibnamefont
  {Carlen}}\ and\ \bibinfo {author} {\bibfnamefont {J.}~\bibnamefont {Maas}},\
  }\href@noop {} {\bibfield  {journal} {\bibinfo  {journal} {Journal of
  Functional Analysis}\ }\textbf {\bibinfo {volume} {273}},\ \bibinfo {pages}
  {1} (\bibinfo {year} {2017})}\BibitemShut {NoStop}%
\bibitem [{\citenamefont {Lindblad}(1976)}]{Lindblad1976}%
  \BibitemOpen
  \bibfield  {author} {\bibinfo {author} {\bibfnamefont {G.}~\bibnamefont
  {Lindblad}},\ }\href {https://doi.org/10.1007/BF01608499} {\bibfield
  {journal} {\bibinfo  {journal} {Communications in Mathematical Physics}\
  }\textbf {\bibinfo {volume} {48}},\ \bibinfo {pages} {119} (\bibinfo {year}
  {1976})}\BibitemShut {NoStop}%
\bibitem [{\citenamefont {Gorini}\  {et~al.}(1976)\citenamefont {Gorini},
  \citenamefont {Kossakowski},\ and\ \citenamefont {Sudarshan}}]{Gorini1976}%
  \BibitemOpen
  \bibfield  {author} {\bibinfo {author} {\bibfnamefont {V.}~\bibnamefont
  {Gorini}}, \bibinfo {author} {\bibfnamefont {A.}~\bibnamefont
  {Kossakowski}},\ and\ \bibinfo {author} {\bibfnamefont {E.~C.~G.}\
  \bibnamefont {Sudarshan}},\ }\href@noop {} {\bibfield  {journal} {\bibinfo
  {journal} {Journal of Mathematical Physics}\ }\textbf {\bibinfo {volume}
  {17}},\ \bibinfo {pages} {821} (\bibinfo {year} {1976})}\BibitemShut
  {NoStop}%
\bibitem [{\citenamefont {Tarasov}(2008)}]{Tarasov2008}%
  \BibitemOpen
  \bibfield  {author} {\bibinfo {author} {\bibfnamefont {V.}~\bibnamefont
  {Tarasov}},\ }\href@noop {} { {\bibinfo {title} {Quantum mechanics of
  non-hamiltonian and dissipative systems}}}\ (\bibinfo  {publisher}
  {Elsevier},\ \bibinfo {year} {2008})\BibitemShut {NoStop}%
\bibitem [{\citenamefont {Nicacio}\  {et~al.}(2016)\citenamefont
  {Nicacio}, \citenamefont {Paternostro},\ and\ \citenamefont
  {Ferraro}}]{Nicacio2016}%
  \BibitemOpen
  \bibfield  {author} {\bibinfo {author} {\bibfnamefont {F.}~\bibnamefont
  {Nicacio}}, \bibinfo {author} {\bibfnamefont {M.}~\bibnamefont
  {Paternostro}},\ and\ \bibinfo {author} {\bibfnamefont {A.}~\bibnamefont
  {Ferraro}},\ }\href {https://doi.org/10.1103/PhysRevA.94.052129} {\bibfield
  {journal} {\bibinfo  {journal} {Physical Review A}\ }\textbf {\bibinfo
  {volume} {94}},\ \bibinfo {pages} {052129} (\bibinfo {year}
  {2016})}\BibitemShut {NoStop}%
\bibitem [{\citenamefont {{Ozorio de A}lmeida}(1998)}]{Ozorio1998}%
  \BibitemOpen
  \bibfield  {author} {\bibinfo {author} {\bibfnamefont {A.~M.}\ \bibnamefont
  {{Ozorio de A}lmeida}},\ }\href
  {https://doi.org/10.1016/S0370-1573(97)00070-7} {\bibfield  {journal}
  {\bibinfo  {journal} {Physics Reports}\ }\textbf {\bibinfo {volume} {295}},\
  \bibinfo {pages} {265} (\bibinfo {year} {1998})}\BibitemShut {NoStop}%
\bibitem [{\citenamefont {Nicacio}\  {et~al.}(2010)\citenamefont
  {Nicacio}, \citenamefont {Maia}, \citenamefont {F.Toscano},\ and\
  \citenamefont {Vallejos}}]{Nicacio2010}%
  \BibitemOpen
  \bibfield  {author} {\bibinfo {author} {\bibfnamefont {F.}~\bibnamefont
  {Nicacio}}, \bibinfo {author} {\bibfnamefont {R.~N.~P.}\ \bibnamefont
  {Maia}}, \bibinfo {author} {\bibnamefont {F.Toscano}},\ and\ \bibinfo
  {author} {\bibfnamefont {R.~O.}\ \bibnamefont {Vallejos}},\ }\href
  {https://doi.org/10.1016/j.physleta.2010.08.076} {\bibfield  {journal}
  {\bibinfo  {journal} {Physics Letters A}\ }\textbf {\bibinfo {volume}
  {374}},\ \bibinfo {pages} {4385} (\bibinfo {year} {2010})}\BibitemShut
  {NoStop}%
\bibitem [{\citenamefont {Frigerio}(1977)}]{Frigerio1977}%
  \BibitemOpen
  \bibfield  {author} {\bibinfo {author} {\bibfnamefont {A.}~\bibnamefont
  {Frigerio}},\ }\href {https://doi.org/10.1007/BF00398571} {\bibfield
  {journal} {\bibinfo  {journal} {Letters in Mathematical Physics}\ }\textbf
  {\bibinfo {volume} {2}},\ \bibinfo {pages} {79–87} (\bibinfo {year}
  {1977})}\BibitemShut {NoStop}%
\bibitem [{\citenamefont {Frigerio}(1978)}]{Frigerio1978}%
  \BibitemOpen
  \bibfield  {author} {\bibinfo {author} {\bibfnamefont {A.}~\bibnamefont
  {Frigerio}},\ }\href {https://doi.org/10.1007/BF01196936} {\bibfield
  {journal} {\bibinfo  {journal} {Communications in Mathematical Physics}\
  }\textbf {\bibinfo {volume} {63}},\ \bibinfo {pages} {269–276} (\bibinfo
  {year} {1978})}\BibitemShut {NoStop}%
\bibitem [{\citenamefont {Carmichael}(1999)}]{Carmichael1999}%
  \BibitemOpen
  \bibfield  {author} {\bibinfo {author} {\bibfnamefont {J.~H.}\ \bibnamefont
  {Carmichael}},\ }\href@noop {} { {\bibinfo {title} {Statistical Methods
  in Quantum Optics 1, Master Equations and {F}okker-{P}lank Equations}}}\
  (\bibinfo  {publisher} {Springer-Verlag},\ \bibinfo {year}
  {1999})\BibitemShut {NoStop}%
\bibitem [{\citenamefont {Alicki}(1976)}]{Alicki1976}%
  \BibitemOpen
  \bibfield  {author} {\bibinfo {author} {\bibfnamefont {R.}~\bibnamefont
  {Alicki}},\ }\href {https://doi.org/10.1016/0034-4877(76)90046-X} {\bibfield
  {journal} {\bibinfo  {journal} {Reports on Mathematical Physics}\ }\textbf
  {\bibinfo {volume} {10}},\ \bibinfo {pages} {249–258} (\bibinfo {year}
  {1976})}\BibitemShut {NoStop}%
\bibitem [{\citenamefont {Gardiner}\ and\ \citenamefont
  {Zoller}(2000)}]{GardinerZoller2000}%
  \BibitemOpen
  \bibfield  {author} {\bibinfo {author} {\bibfnamefont {C.~W.}\ \bibnamefont
  {Gardiner}}\ and\ \bibinfo {author} {\bibfnamefont {P.}~\bibnamefont
  {Zoller}},\ }\href@noop {} { {\bibinfo {title} {Quantum noise}}},\
  \bibinfo {edition} {2nd}\ ed.,\ Springer series in synergetics\ (\bibinfo
  {publisher} {Springer},\ \bibinfo {address} {Berlin ; Heidelberg [u.a.]},\
  \bibinfo {year} {2000})\BibitemShut {NoStop}%
\bibitem [{\citenamefont {Breuer}\  {et~al.}(2002)\citenamefont {Breuer},
  \citenamefont {Breuer}, \citenamefont {Petruccione},\ and\ \citenamefont
  {Petruccione}}]{Breuer2002}%
  \BibitemOpen
  \bibfield  {author} {\bibinfo {author} {\bibfnamefont {H.}~\bibnamefont
  {Breuer}}, \bibinfo {author} {\bibfnamefont {P.}~\bibnamefont {Breuer}},
  \bibinfo {author} {\bibfnamefont {F.}~\bibnamefont {Petruccione}},\ and\
  \bibinfo {author} {\bibfnamefont {S.}~\bibnamefont {Petruccione}},\ }\href
  {https://books.google.com.br/books?id=0Yx5VzaMYm8C} { {\bibinfo {title}
  {The Theory of Open Quantum Systems}}}\ (\bibinfo  {publisher} {Oxford
  University Press},\ \bibinfo {year} {2002})\BibitemShut {NoStop}%
\bibitem [{\citenamefont {Wiseman}\ and\ \citenamefont
  {Milburn}(2009)}]{WisemanMilburn2009}%
  \BibitemOpen
  \bibfield  {author} {\bibinfo {author} {\bibfnamefont {H.~M.}\ \bibnamefont
  {Wiseman}}\ and\ \bibinfo {author} {\bibfnamefont {G.~J.}\ \bibnamefont
  {Milburn}},\ }\href {https://doi.org/10.1017/CBO9780511813948} {
  {\bibinfo {title} {Quantum Measurement and Control}}}\ (\bibinfo  {publisher}
  {Cambridge University Press},\ \bibinfo {year} {2009})\BibitemShut {NoStop}%
\bibitem [{\citenamefont {Banchi}\  {et~al.}(2015)\citenamefont {Banchi},
  \citenamefont {Braunstein},\ and\ \citenamefont {Pirandola}}]{Banchi2015}%
  \BibitemOpen
  \bibfield  {author} {\bibinfo {author} {\bibfnamefont {L.}~\bibnamefont
  {Banchi}}, \bibinfo {author} {\bibfnamefont {S.~L.}\ \bibnamefont
  {Braunstein}},\ and\ \bibinfo {author} {\bibfnamefont {S.}~\bibnamefont
  {Pirandola}},\ }\href {https://doi.org/10.1103/PhysRevLett.115.260501}
  {\bibfield  {journal} {\bibinfo  {journal} {Physical Review Letters}\
  }\textbf {\bibinfo {volume} {115}},\ \bibinfo {pages} {260501} (\bibinfo
  {year} {2015})}\BibitemShut {NoStop}%
\bibitem [{\citenamefont {Simon}\  {et~al.}(1994)\citenamefont {Simon},
  \citenamefont {Mukunda},\ and\ \citenamefont {Dutta}}]{Simon1994}%
  \BibitemOpen
  \bibfield  {author} {\bibinfo {author} {\bibfnamefont {R.}~\bibnamefont
  {Simon}}, \bibinfo {author} {\bibfnamefont {N.}~\bibnamefont {Mukunda}},\
  and\ \bibinfo {author} {\bibfnamefont {B.}~\bibnamefont {Dutta}},\ }\href
  {https://doi.org/10.1103/PhysRevA.49.1567} {\bibfield  {journal} {\bibinfo
  {journal} {Phys. Rev. A}\ }\textbf {\bibinfo {volume} {49}},\ \bibinfo
  {pages} {1567} (\bibinfo {year} {1994})}\BibitemShut {NoStop}%
\bibitem [{\citenamefont {Nicacio}(2021{\natexlab{a}})}]{Nicacio2021wil}%
  \BibitemOpen
  \bibfield  {author} {\bibinfo {author} {\bibfnamefont {F.}~\bibnamefont
  {Nicacio}},\ }\href {https://doi.org/10.1119/10.0005944} {\bibfield
  {journal} {\bibinfo  {journal} {American Journal of Physics}\ }\textbf
  {\bibinfo {volume} {89}},\ \bibinfo {pages} {1139} (\bibinfo {year}
  {2021}{\natexlab{a}})},\ \Eprint
  {https://arxiv.org/abs/https://doi.org/10.1119/10.0005944}
  {https://doi.org/10.1119/10.0005944} \BibitemShut {NoStop}%
\bibitem [{\citenamefont {Nicacio}(2021{\natexlab{b}})}]{Nicacio2021Weyl}%
  \BibitemOpen
  \bibfield  {author} {\bibinfo {author} {\bibfnamefont {F.}~\bibnamefont
  {Nicacio}},\ }\href {https://doi.org/10.1088/1751-8121/abd5c6} {\bibfield
  {journal} {\bibinfo  {journal} {Journal of Physics A: Mathematical and
  Theoretical}\ }\textbf {\bibinfo {volume} {54}},\ \bibinfo {pages} {055004}
  (\bibinfo {year} {2021}{\natexlab{b}})}\BibitemShut {NoStop}%
\bibitem [{\citenamefont {de~Gosson}(2006)}]{deGosson2006}%
  \BibitemOpen
  \bibfield  {author} {\bibinfo {author} {\bibfnamefont {M.}~\bibnamefont
  {de~Gosson}},\ }\href {https://books.google.com.br/books?id=q9SHRvay75IC}
  { {\bibinfo {title} {Symplectic Geometry and Quantum Mechanics}}},\
  Operator Theory: Advances and Applications\ (\bibinfo  {publisher}
  {Birkh{\"a}user Basel},\ \bibinfo {year} {2006})\BibitemShut {NoStop}%
\bibitem [{\citenamefont {Risken}(1996)}]{Risken1996}%
  \BibitemOpen
  \bibfield  {author} {\bibinfo {author} {\bibfnamefont {H.}~\bibnamefont
  {Risken}},\ }\href@noop {} { {\bibinfo {title} {The {F}okker-{P}lanck
  Equation: Methods of Solution and Applications}}}\ (\bibinfo  {publisher}
  {Springer},\ \bibinfo {year} {1996})\BibitemShut {NoStop}%
\bibitem [{\citenamefont {Horn}\ and\ \citenamefont
  {Johnson}(1991)}]{Horn1991}%
  \BibitemOpen
  \bibfield  {author} {\bibinfo {author} {\bibfnamefont {R.~A.}\ \bibnamefont
  {Horn}}\ and\ \bibinfo {author} {\bibfnamefont {C.~R.}\ \bibnamefont
  {Johnson}},\ }\href {https://doi.org/10.1017/CBO9780511840371} {
  {\bibinfo {title} {Topics in Matrix Analysis}}}\ (\bibinfo  {publisher}
  {Cambridge University Press},\ \bibinfo {year} {1991})\BibitemShut {NoStop}%
\bibitem [{\citenamefont {Valero-Toranzo}\  {et~al.}(2018)\citenamefont
  {Valero-Toranzo}, \citenamefont {Zozor},\ and\ \citenamefont
  {Brossier}}]{Toranzo2017}%
  \BibitemOpen
  \bibfield  {author} {\bibinfo {author} {\bibfnamefont {I.}~\bibnamefont
  {Valero-Toranzo}}, \bibinfo {author} {\bibfnamefont {S.}~\bibnamefont
  {Zozor}},\ and\ \bibinfo {author} {\bibfnamefont {J.}~\bibnamefont
  {Brossier}},\ }\href {https://doi.org/10.1109/TIT.2017.2771209} {\bibfield
  {journal} {\bibinfo  {journal} {IEEE Transactions on Information Theory}\
  }\textbf {\bibinfo {volume} {64}},\ \bibinfo {pages} {6743} (\bibinfo {year}
  {2018})}\BibitemShut {NoStop}%
\bibitem [{\citenamefont {Bialas}\ and\ \citenamefont
  {Gora}(2015)}]{Bialas2015}%
  \BibitemOpen
  \bibfield  {author} {\bibinfo {author} {\bibfnamefont {S.}~\bibnamefont
  {Bialas}}\ and\ \bibinfo {author} {\bibfnamefont {M.}~\bibnamefont {Gora}},\
  }\href {https://doi.org/10.1515/bpasts-2015-0018} {\bibfield  {journal}
  {\bibinfo  {journal} {Bulletin of the Polish Academy of Sciences Technical
  Sciences}\ }\textbf {\bibinfo {volume} {63}},\ \bibinfo {pages} {163–168}
  (\bibinfo {year} {2015})}\BibitemShut {NoStop}%
\bibitem [{\citenamefont {Fassbender}\  {et~al.}(1999)\citenamefont
  {Fassbender}, \citenamefont {Mackey}, \citenamefont {Mackey},\ and\
  \citenamefont {Xu}}]{Fassbender1999}%
  \BibitemOpen
  \bibfield  {author} {\bibinfo {author} {\bibfnamefont {H.}~\bibnamefont
  {Fassbender}}, \bibinfo {author} {\bibfnamefont {D.~S.}\ \bibnamefont
  {Mackey}}, \bibinfo {author} {\bibfnamefont {N.}~\bibnamefont {Mackey}},\
  and\ \bibinfo {author} {\bibfnamefont {H.}~\bibnamefont {Xu}},\ }\href
  {https://doi.org/10.1016/S0024-3795(98)10137-4} {\bibfield  {journal}
  {\bibinfo  {journal} {Linear Algebra and its Applications}\ }\textbf
  {\bibinfo {volume} {287}},\ \bibinfo {pages} {125–159} (\bibinfo {year}
  {1999})}\BibitemShut {NoStop}%
\bibitem [{\citenamefont {Gardiner}\ and\ \citenamefont
  {Zoller}(2004)}]{GardinerZoller2004}%
  \BibitemOpen
  \bibfield  {author} {\bibinfo {author} {\bibfnamefont {C.~W.}\ \bibnamefont
  {Gardiner}}\ and\ \bibinfo {author} {\bibfnamefont {P.}~\bibnamefont
  {Zoller}},\ }\href@noop {} { {\bibinfo {title} {Quantum noise : a
  handbook of Markovian and non-Markovian quantum stochastic methods with
  applications to quantum optics}}},\ \bibinfo {edition} {3rd}\ ed.\ (\bibinfo
  {publisher} {Springer},\ \bibinfo {year} {2004})\BibitemShut {NoStop}%
\bibitem [{\citenamefont {Toscano}\  {et~al.}(2005)\citenamefont
  {Toscano}, \citenamefont {de~Matos~Filho},\ and\ \citenamefont
  {Davidovich}}]{Toscano2005}%
  \BibitemOpen
  \bibfield  {author} {\bibinfo {author} {\bibfnamefont {F.}~\bibnamefont
  {Toscano}}, \bibinfo {author} {\bibfnamefont {R.~L.}\ \bibnamefont
  {de~Matos~Filho}},\ and\ \bibinfo {author} {\bibfnamefont {L.}~\bibnamefont
  {Davidovich}},\ }\href {https://doi.org/10.1103/PhysRevA.71.010101}
  {\bibfield  {journal} {\bibinfo  {journal} {Physical Review A}\ }\textbf
  {\bibinfo {volume} {71}},\ \bibinfo {pages} {010101(R)} (\bibinfo {year}
  {2005})}\BibitemShut {NoStop}%
\bibitem [{\citenamefont {Horn}\ and\ \citenamefont
  {Johnson}(1990)}]{Horn1990}%
  \BibitemOpen
  \bibfield  {author} {\bibinfo {author} {\bibfnamefont {R.~A.}\ \bibnamefont
  {Horn}}\ and\ \bibinfo {author} {\bibfnamefont {C.~R.}\ \bibnamefont
  {Johnson}},\ }
  { {\bibinfo {title} {Topics in Matrix Analysis}}}\ (\bibinfo  {publisher}
  {Cambridge University Press, New York},\ \bibinfo {year} {1994})\BibitemShut {NoStop}%
\bibitem [{\citenamefont {Horn}\ and\ \citenamefont
  {Johnson}(2013)}]{Horn2013}%
  \BibitemOpen
  \bibfield  {author} {\bibinfo {author} {\bibfnamefont {R.~A.}\ \bibnamefont
  {Horn}}\ and\ \bibinfo {author} {\bibfnamefont {C.~R.}\ \bibnamefont
  {Johnson}},\ }\href@noop {} { {\bibinfo {title} {Matrix Analysis}}},\
  \bibinfo {edition} {2nd}\ ed.\ (\bibinfo  {publisher} {Cambridge University
  Press},\ \bibinfo {address} {Cambridge; New York},\ \bibinfo {year}
  {2013})\BibitemShut {NoStop}%
\end{thebibliography}

%

\end{document}